\documentclass[prc,showpacs,showkeys]{revtex4}
\usepackage{epsfig,graphicx,float,color}
\usepackage{amssymb}
\usepackage[normalem]{ulem}

\begin{document}

\title{Nuclear shape coexistence: A study of the even-even Hg isotopes
  using the interacting  boson model with configuration mixing}

\author{J.E.~Garc\'{\i}a-Ramos$^1$ and K.~Heyde$^2$}
\affiliation{
$^1$Departamento de  F\'{\i}sica Aplicada, Universidad de Huelva,
21071 Huelva, Spain\\
$^2$Department of Physics and Astronomy, Ghent University,
Proeftuinstraat, 86 B-9000 Gent, Belgium}

\begin{abstract} 
\begin{description}
\item [Background:] The Po, Pb, Hg, and Pt region is known for the presence of
coexisting structures that correspond to different particle-hole
configurations in the Shell Model language or equivalently to nuclear shapes
with different deformation.   

\item [Purpose:] We intend to study the configuration mixing phenomenon
in the Hg isotopes and to understand how different observables are influenced by it.
 
\item [Method:]  We study in detail a long chain of mercury isotopes,
$^{172-200}$Hg, using the interacting boson model with configuration
mixing. The parameters of the Hamiltonians are fixed through a least
square fit to the known energies and absolute B(E2) transition rates
of states up to $3$ MeV.

\item [Results:] We obtained the IBM-CM Hamiltonians and we calculate
excitation energies, B(E2)'s, quadrupole 
shape invariants, wave functions, isotopic shifts, and mean field
energy surfaces.  
 
\item [Conclusions:] We obtain a fairly good agreement with the experimental
data for all the studied observables and we conclude that the Hamiltonian
and the states we obtain constitute a good approximation to the Hg
isotopes. 
\end{description}
\end{abstract}


\maketitle

\section{Introduction}
\label{sec-intro}

The exploration of the many facets of the atomic nucleus over many decades, using a large
set of complementary probes, using in particular the electromagnetic (nuclear electric and
magnetic transition probabilities and moments, mapping of nuclear charge radii over large
series of isotopes) and strong forces, has unambiguously shown the appearance of the essential degrees
of freedom. Both, few nucleon properties (near closed shells and doubly-closed shells) as
well as collective properties have been discovered \cite{bomot75}, most often going hand in hand with the
increasing technical possibilities that bring even nuclei far from the region of $\beta$ stability
within reach (see in particular Chapter 1 of Rowe and Wood \cite{rowe10}).

A theoretical description of the atomic nucleus, viewed as a system of A nucleons
(Z protons and N neutrons) interacting through an in-medium effective nucleon force has
reached important progress during the last decades \cite{epel09,mach11,caurier05,bender03}. 
However, it looks like generic characteristics
of nuclear excitations stem from the interplay of, on one side, the low multipoles of that interaction
that generate nuclear mean-field properties, characterized by a nuclear shell structure
and, on the other side, the high multipoles, scattering the interacting nucleons out of their independent
particle orbitals, evidenced by the nuclear pairing properties characterized by an energy gap that
allows nuclear superfluidity \cite{dean03} to appear along long series of isotopes and isotones. 

It seems that the balance between these two opposing nuclear force components, \textit{i.e.}, on one side
the stabilizing effect of closed shells which prefers nuclei to exhibit a spherical shape, versus
a redistribution of protons and/or neutrons in a more deformed shape is at the origin of the
appearance of nuclear shape coexistence. By now, nuclear shape coexistence has evolved from the
early 
interpretation of Morinaga \cite{morinaga56}, into a phenomenon that appears all through the nuclear
landscape, both in light nuclei (near the N $=8,20,28,40$ neutron closed shells) as well as in heavy
nuclei (see \cite{jul01,neyens03,blaum13,gade08,gorgen10,bauman12,blum13} and \cite{hey83,wood92,heyde11} for an extensive discussion of 
both the experimental methods and theoretical model approaches over a period of about 3 decades).

Two naturally complementary roads can be taken in order to describe
the phenomenon of nuclear shape coexistence. 
Starting from a nuclear shell-model approach, protons and neutrons are
expected to gradually fill the various 
shells at Z, N $=2,8,20,28,$ ... giving rise to a number of
double-closed shell nuclei that are the reference 
points determining shells in which valence nucleons have been allowed to interact through either a
phenomenologically fitted effective interaction or a microscopic
effective interaction, deduced from many-body 
theory from realistic NN forces \cite{barrett13,bogner10,cora09}. 
In view of the large evidence that multi-particle multi-hole excitations 
are observed, even at a rather low excitation energy, it is important to delineate those regions in the
nuclear mass table where conditions are such that shape coexistence may show up. It turns out that it is
the balance between the cost to excite such mp-nh excitations at first and the energy gain that results
from the enlarged availability of protons and neutrons to interact strongly through the low nuclear multipoles
such as to give rise to a ``deformed-spherical'' inversion or the presence of low-lying competing shape coexisting
configurations. With the advent of a strongly increased computing power as well as of the construction of improved
algorithms to determine the energy eigenvalues of very big energy matrices, a very large body of calculations mainly 
along series of isotones at N $=8,20, 28, 40,$ 
and, very recently, at N=50 has appeared in the literature. Even the
well-known doubly-closed shell nuclei  
$^{16}$O, $^{40}$Ca, $^{56}$Ni, $^{48}$Ca, ... exhibit a number of
mp-mh excitations. In these calculations, it is 
paramount to treat the change in the monopole part of the nuclear
field (the changing single-particle energies) 
as well as the other multipoles originating from the nuclear
interaction in a consistent way \cite{sorlin08,otsuka13}. 

In the other approach, the starting point is an effective nuclear
force or energy-density functional which are 
used to derive the optimized single-particle basis in a
self-consistent way, making use of Hartree-Fock (HF), or using 
Hartree-Fock-Bogoliubov (HFB) theory, when also including
the strong nucleon pairing forces in both cases constraining the nuclear
density distribution to specific values for the quadrupole moments,
octupole moments, etc \cite{bender03,drut10}.  
In order to confront the results of mean-field studies with the
experimental data, one needs to restore the symmetries that are 
broken in the construction of a HF or HFB reference state. It is necessary therefore to construct states that correspond
to a fixed proton (Z) and neutron number (N) as well as good angular momentum J (including K for deformed nuclei). These
states then form the starting basis to introduce the dynamical collective correlations beyond the mean-field energy (solving
the Hill-Wheeler equations). This gives rise to energy spectra and many other properties such as B(E2) values, quadrupole moments,
charge radii and E0 transition rates and can serve as a very good basis to be confronted with the data. Besides the use of various
effective forces in the HF (HFB) approach (Skyrme force \cite{erler11}, Gogny-force \cite{gogny80,gogny88}), a 
relativistic mean-field approach has been formulated early
on by Walecka \cite{walecka74} and improved over the years into a microscopic effective-field theory 
\cite{serot86,reinhard89,serot92,ring96,niksic11}.
It has been shown that the Generator Coordinate Method (GCM) with the Gaussian overlap approximation (GOA) reduces the problem of 
solving the Hill-Wheeler equations into solving an equivalent collective Bohr Hamiltonian for the full five-dimensional
collective model \cite{bender03}. This approach has been used frequently with Gogny forces
within the framework of both standard constrained Hartree-Fock-Bogoliubov (CHFB) calculations~\cite{dela10}  as well as making use of relativistic 
mean-field methods~\cite{niksic11}.

A particularly well-documented example of shape coexistence shows up in the Pb region. From the closed neutron
shell (N=126) to the very neutron-deficient nuclei, approaching and
even going beyond the N=104 mid-shell nuclei, ample experimental evidence
for shape coexisting bands has accumulated for the Pb (Z=82),
the Po (Z=84), the
Hg (Z=80) and the Pt (Z=78) nuclei \cite{jul01,hey83,wood92,heyde11}. Major steps have been taken over a period of more than 3 decades
since the early work, disclosing the presence of low-lying 0$^+$ states in the $^{192-198}$Pb nuclei \cite{duppen84}. 
The discovery of three shape coexisting configurations in the mid-shell $^{186}$Pb nucleus (at N=104) \cite{andrei00}
even accelerated the accumulation of new data since 2010.
Very recent 
experimental campaigns have largely extended our knowledge beyond the excitation energies of intruding bands by providing
information on nuclear lifetimes \cite{dewald03,grahn06,grahn08,grahn09,scheck10}, nuclear charge radii
\cite{dewitte07,cocio11,seliv13}, 2$_1^+$ gyromagnetic factors
\cite{stuchbery96,bian07}, $\alpha$-decay hindrance factors
\cite{wauters94,wauters94a,richards96,uusitalo97,duppen00,delion95,richards97} and, very recently, Coulomb excitation 
using radioactive beams at REX-ISOLDE (CERN) \cite{Isol12}.  

An important question is how these shape coexisting structures will evolve when one moves further away from 
the Z=82 and N=126 closed shells. 
Whereas the intruder bands are easily singled out for the Pb and Hg nuclei in which 
the excitation energies display the characteristic parabolic pattern with
minimal excitation energy around the N=104 neutron mid-shell nucleus, this
structure is not immediate in both the Pt and the Po nuclei.

Theoretical efforts have been carried out over the years, both
exploring the mean-field behaviour, even going beyond including the
collective dynamics, as 
well as making use of symmetry-dictated truncated shell-model calculations. 

Early calculations started from a deformed Woods-Saxon potential, 
exploring the nuclear energy surfaces as a function of the quadrupole deformation
variables \cite{may77,bengt87,bengt89,naza93} and showed a consistent
picture pointing 
to the presence of oblate and prolate energy minima when moving away
from the double-closed shell $^{208}$Pb nucleus.
More recently, HF and HFB mean-field calculations going beyond the static part,
including dynamical effects through the use of the Generator Coordinate Method
(GCM) \cite{bender03}, either starting from Skyrme functionals
\cite{duguet03,smirnova03,bender04,grahn08,yao13}, or using the Gogny D1S
parameterization \cite{girod89,dela94,chasman01,egido04,rodri04,sarri08,rodri10} have put phenomenological
calculations on a firm ground, moreover giving rise to
detailed information concerning the collective bands observed in
neutron-deficient nuclei around the Z=82 closed proton shell. Calculations in order to understand possible shape 
changes and shape transitions in the Pb region 
within the relativistic mean field (RMF) approach \cite{sharma92,patra94,yoshida94,yoshida97,fossion06,niksic02,niksic10,
niksic11} have been improved with increasing sophistication. 

From a microscopic shell-model point of view, the hope to treat on equal footing the
large open neutron shell from N=126 down to and beyond the mid-shell
N=104 region, jointly with the valence protons in the Pt, Hg, Po, and Rn
nuclei, even including proton multi-particle multi-hole (mp-nh) excitations across
the Z=82 shell closure, is far beyond present computational possibilities.
The truncation of the model space, however, by concentrating on nucleon pair
modes (mainly $0^+$ and $2^+$ coupled pairs, to be treated as bosons
within the interacting boson approximation (IBM) \cite{iach87}),
has made the calculations feasible, even including pair excitations
across the Z=82 shell closure \cite{duval81,duval82} in the Pb region in a transparent way.
More in particular, the Pb nuclei have been extensively studied giving rise
to bands with varying collectivity depending on the nature of the
excitations treated in the model space
\cite{hey87,hey91,fossion03,helle05,paka07,helle08}.
More recently, detailed studies of the Pt nuclei have been carried out \cite{king98,harder97,Garc09,Garc11} as 
an attempt to describe the larger amount of the low-lying states and their
E2 decay properties 
explicitly including particle-hole excitations across the Z=82 shell
closure. However, in other studies \cite{cutcham05a,cutcham05} the Pt nuclei
were treated without the inclusion of such particle-hole excitations. 

In the present paper, an extensive study of the even-even Hg is
carried 
explicitly including particle-hole excitations across the Z=82 shell closure.
More specifically, within the IBM configuration-mixing approach \cite{duval81,duval82}
(IBM-CM for short), early calculations were carried out for the Hg nuclei in the mass region
182$\leq$A$\leq$192, with more extensive studies extending the mass region up to A=202 \cite{bar83,bar84}. 
The region 194$\leq$A$\leq$202, which exhibits no indication of intruder states was subsequently
studied by Druce \textit{et al.} \cite{druce87}. These studies made use of the proton-neutron formulation
of the configuration mixing IBM. Since in the Pb, Hg, and Pt nuclei, one expects the lowest-lying excited
states to be described by the fully symmetric configurations, the concept of maximal F-spin \cite{arima77,otsuka78,barrett91} 
can be used 
and allows for the possibility to no longer discriminate between proton and neutron bosons. In the present
mass region, however, the intruder excitations do play a dominant role and influence to a large extent
the observed structures in these isotopes (as has been shown to be the case for the Pb(\cite{helle05,helle08}) 
and the Pt (\cite{harder97,king98,Garc09,Garc11}) nuclei before).
The interacting boson model has been used and applied in the Pb region, in particular concentrating on shape
coexistence in the Pb region making use of a totally different approach. The parameters of the IBM Hamiltonian
are determined from mapping the total energy surface, derived from self-consistent HFB calculations using
the Gogny D1S and D1M energy functionals \cite{nomura08,nomura10} onto the corresponding IBM mean-field energy
\cite{gino80,diep80a,diep80b}. In particular the Pt isotopes \cite{nomura11a,nomura11b}, the Hg isotopes
\cite{nomura13} and the Pb isotopes \cite{nomura12a} have been studied.

\begin{figure}[hbt]
  \centering
  \includegraphics[width=0.90\linewidth]{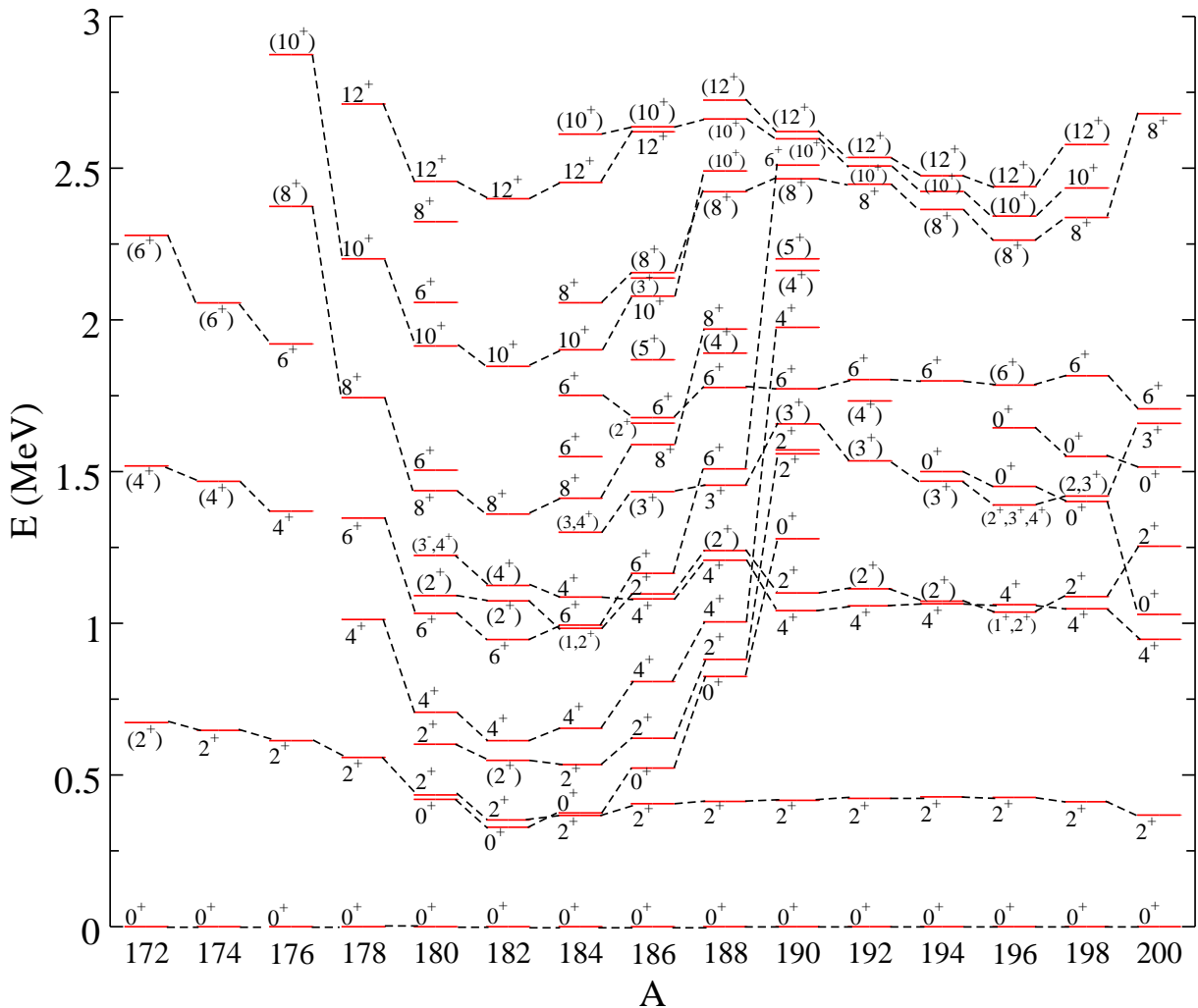}
\caption{(Color online) Experimental energy level systematics for Hg isotopes.
  Only levels up to $E_x \sim$ 3.0 MeV are shown.}
  \label{fig-system-hg}
\end{figure}

\section{The experimental data in the even-even Hg nuclei}
\label{sec-expe}

The even-even Hg nuclei span a very large region of isotopes, starting with the lightest presently know $^{172}$Hg nucleus (N=92),
passing through the mid-shell point at N=104 at $^{184}$Hg, all the way up to the N=126 neutron closed shell at $^{206}$Hg.
Many experimental complementary methods have been used to disentangle the properties over such a large interval. These nuclei
are extensively covered in the Nuclear Data Sheet reviews for A=172 \cite{Kibedi10}, A=174 \cite{Browne99}, A=176 \cite{Basu06},
A=178 \cite{Achter09}, A=180 \cite{Wu03}, A=182 \cite{Sing11}, A=184 \cite{Bagl10}, A=186 \cite{Bagl03}, A=188 \cite{Sing02},
A=190 \cite{Sing03}, A=192 \cite{Bagl12}, A=194 \cite{Sing06}, A=196 \cite{Xia07}, A=198 \cite{Xia02}, and A=200 \cite{Kondev07},
and span the region we concentrate on in the present paper. Moreover, we have incorporated the most recent papers (up to and 
including early 2013) in order to highlight the salient experimental features of the Hg nuclei,over the mass span from A=172 to mass A=200.

The yrast band structure for mass A=172 to A=176 have been studied using the highly-selective recoil decay tagging (RDT) technique
\cite{sand09,jul01,muikku98}. Carpenter \textit{et al.} \cite{carpent97} studied both the A=176 and A=178 Hg nuclei. Using 
Gammasphere at the Fragment Mass Analyzer (FMA), the yrast band structure could be considerably extended \cite{kondev99,kondev00} for both mass A=178 and
A=180. Experiments, in the late 80's \cite{simon86,drac88} showed hints of shape coexisting states. Recent developments
in the experimental methods to study the nuclear structure properties of these neutron-deficient Hg nuclei, allowed to substantially
increase the nuclear structure data basis: ($\beta^+$/EC)-decay of $^{180}$Tl \cite{else11} and in-beam conversion-electron
spectroscopy \cite{page11}.  Besides the new information on the low-spin states below E$_x \sim$ 2 MeV, lifetimes up to spin
J$^{\pi}$ = 8${^+}$(10$^+$) have been measured from using Recoil-Decay tagged (RDT) $\gamma$-rays, using the recoil distance Doppler-shift 
(RDDS) method 
\cite{grahn09,scheck10} for mass A=180 and A=182. This recent information largely extends the early data (see refs. 
\cite{bindra95,wauters96,ma84}). 
Except for mass A=184 (new results on E0 transitions from conversion electron and $\gamma$-ray studies~\cite{scheck11}), no new results have
been obtained since the most recent NDS evaluations, as cited before.

For the mass A=196 and A=198 Hg isotopes, recent experiments using the HORUS cube $\gamma$-ray spectrometer and  
$\gamma-\gamma$ angular correlation measurements, made it possible to
determine multipole mixing ratios, spins and energy 
spectra \cite{bern10,bern09}. Moreover, in the case of A=198, (p,t) two-neutron transfer reactions allowed to map out the 
presence of a number of 0$^+$ states \cite{bern13}.

The experimental energy systematics derived from the above data set for the Hg nuclei is shown in Fig.~\ref{fig-system-hg} 
and spans the interval A=172 up to and including A=200. The systematics is limited (for the high-spin states) up to E$_x \sim$
3 MeV. For mass numbers A $\geq$ 190, above the energy of E$_x \sim$ 1.5 MeV, a number of low-spin states without a unique
spin-parity assignment from the NDS evaluations are not drawn (often states with negative parity and/or more spin possibilities). 

In between mass A=178 progressing to the lower A=176-172 isotopes, no connecting dashed lines are drawn because the present
data set does not contain unambiguous information in order to extend
them (see, \textit{e.g.}, the behavior of the J$^{\pi}$=4$^+$  
state going from A=178 to A=176). What is clear though is that from mass A=180 and downwards, the 2$^+_1$ excitation energy
is steadily increasing (as well as the energy of the associated (4$^+$) and (6$^+$) states). The interval 180 $\leq$ A $\leq$ 188(190)
exhibits the presence of a 0$^+$, 2$^+$, 4$^+$, 6$^+$, 8$^+$, 10$^+$
collective band structure with a ``parabolic''-like energy 
dependence on N (with respect to the minimal energy of the 0$^+_2$ state at N=102). For the heavier nuclei (A $>$ 190), the Hg structure
exhibits an almost flat behavior of the excitation energy as a function of increasing mass number A (or neutron number N). The
observed  0$^+_1$, 2$^+_1$, 4$^+_1$, 2$^+_2$,... sequence seems to be pointing out the appearance of a $\gamma$-soft structure.

In the present paper, we cover the whole interval 172 $\leq$ A $\leq$ 200, but in the discussion mainly concentrate on the A=180-188(190) 
region which forms a challenge to theoretical model approaches.
The energy systematics as shown in  Fig.~\ref{fig-system-hg} indicates three distinct structures: a less deformed one for mass A $>$ 190, 
a region where a more collective structure intrudes to low excitation energies in the region 180 $\leq$ A $\leq$ 188, and a region where the 
lowest 2$^+$ state (and associated higher-spin states for the yrast part) exhibits a steadily increasing excitation energy (A $\leq$ 178).    
\begin{figure}[hbt]
  \centering
  \includegraphics[width=0.8\linewidth]{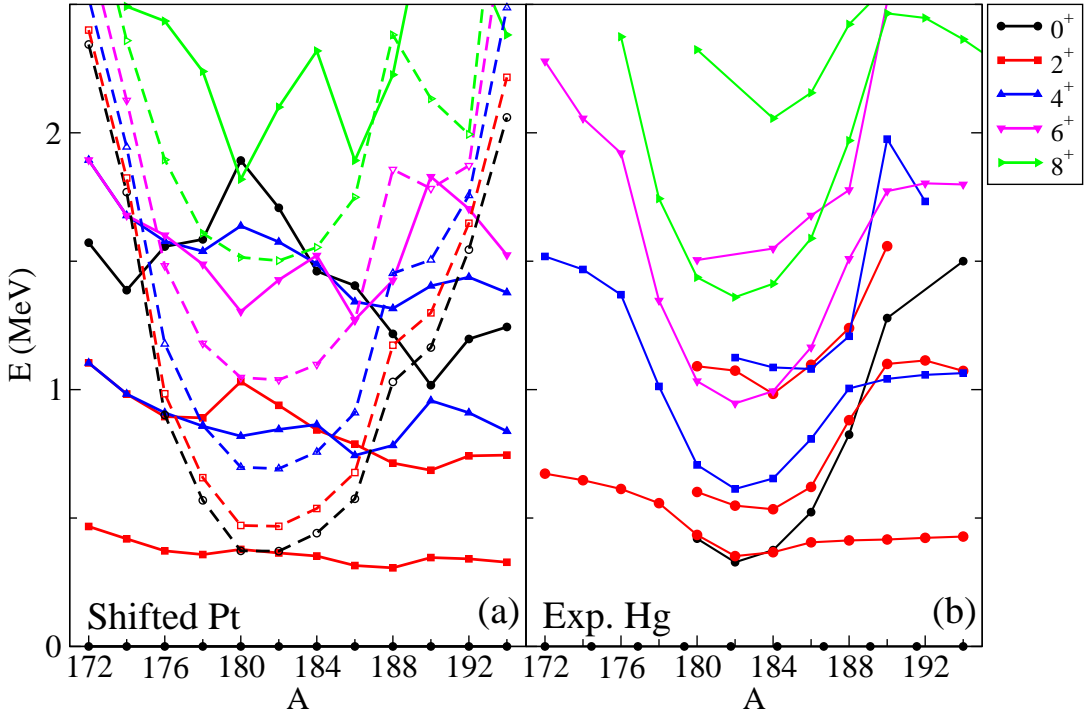}
  \caption{(Color online) Comparison between the ``shifted''
    theoretical energy spectra for Pt (panel (a)) (see text and Figure 12 in ~\cite{Garc09})
    and the experimental Hg low-lying energy spectra (panel (b)).}
  \label{fig-energ-hg}
\end{figure}

\section{The Interacting Boson Model with configuration
 mixing formalism}
\label{sec-ibm-cm}
\subsection{The formalism}
\label{sec-formalism}

The IBM-CM allows the simultaneous treatment and mixing of several
boson configurations which correspond to different particle--hole
(p--h) shell-model excitations \cite{duval82}. On the basis of
intruder spin symmetry \cite{hey92,hey94,coster96,oros99}, no distinction is made
between particle and hole bosons. Hence, the model space which
includes the regular proton 2h configurations and a number of valence
neutrons outside of the N=82 closed shell (corresponding to the
standard IBM treatment for the Hg even-even nuclei) as well as the
proton 4h-2p configurations and the same number of valence neutrons
corresponds to a $[N]\oplus[N+2]$ boson space ($N$ being the number of
active protons, counting both proton holes and particles, plus the
number of valence neutrons outside the N=82 closed shell,
divided by $2$ as the boson number). Consequently, the Hamiltonian for
two configuration mixing can be written as
\begin{equation}
  \hat{H}=\hat{P}^{\dag}_{N}\hat{H}^N_{\rm ecqf}\hat{P}_{N}+
  \hat{P}^{\dag}_{N+2}\left(\hat{H}^{N+2}_{\rm ecqf}+
    \Delta^{N+2}\right)\hat{P}_{N+2}\
  +\hat{V}_{\rm mix}^{N,N+2}~,
\label{eq:ibmhamiltonian}
\end{equation}
where $\hat{P}_{N}$ and $\hat{P}_{N+2}$ are projection operators onto
the $[N]$ and the $[N+2]$ boson spaces, 
respectively, $\hat{V}_{\rm mix}^{N,N+2}$  describes
the mixing between the $[N]$ and the $[N+2]$ boson subspaces, and
\begin{equation}
  \hat{H}^i_{\rm ecqf}=\varepsilon_i \hat{n}_d+\kappa'_i
  \hat{L}\cdot\hat{L}+
  \kappa_i
  \hat{Q}(\chi_i)\cdot\hat{Q}(\chi_i), \label{eq:cqfhamiltonian}
\end{equation}
is the extended consistent-Q Hamiltonian (ECQF) \cite{warner83} with $i=N,N+2$,
$\hat{n}_d$ the $d$ boson number operator, 
\begin{equation}
  \hat{L}_\mu=[d^\dag\times\tilde{d}]^{(1)}_\mu ,
\label{eq:loperator}
\end{equation}
the angular momentum operator, and
\begin{equation}
  \hat{Q}_\mu(\chi_i)=[s^\dag\times\tilde{d}+ d^\dag\times
  s]^{(2)}_\mu+\chi_i[d^\dag\times\tilde{d}]^{(2)}_\mu~,\label{eq:quadrupoleop}
\end{equation}
the quadrupole operator. 
We are not considering the most general
IBM Hamiltonian in each Hilbert space, [N] and [N+2], but we are
restricting 
to a ECQF formalism \cite{warner83,lipas85} in each subspace. This
approach has been shown to be a rather good approximation in many
calculations and in particular in two recent papers describing the Pt isotopes \cite{Garc09,Garc11}. 

The parameter $\Delta^{N+2}$ can be
associated with the energy needed to excite two proton particles across the
Z=82 shell gap, giving rise to 2p-2h excitations, corrected for the pairing interaction gain and including
monopole effects~\cite{hey85,hey87}.
The operator $\hat{V}_{\rm mix}^{N,N+2}$ describes the mixing between
the $N$ and the $N+2$ configurations and is defined as
\begin{equation}
  \hat{V}_{\rm mix}^{N,N+2}=w_0^{N,N+2}(s^\dag\times s^\dag + s\times
  s)+w_2^{N,N+2} (d^\dag\times d^\dag+\tilde{d}\times \tilde{d})^{(0)}.
\label{eq:vmix}
\end{equation}

The $E2$ transition operator for two-configuration mixing is
subsequently defined as
\begin{equation}
  \hat{T}(E2)_\mu=\sum_{i=N,N+2} e_i
  \hat{P}_i^\dag\hat{Q}_\mu(\chi_i)\hat{P}_i~,\label{eq:e2operator}
\end{equation}
where the $e_i$ ($i=N,N+2$) are the effective boson charges and
$\hat{Q}_\mu(\chi_i)$ the quadrupole operator defined in equation
(\ref{eq:quadrupoleop}).

In section \ref{sec-fit-procedure} we present the methods used in order to 
determine the parameters
appearing in the IBM-CM Hamiltonian as well as in the $\hat{T}(E2)$ operator.

The wave function, within the IBM-CM, can be described as 
\begin{eqnarray}
\Psi(k,JM) &=& \sum_{i} a^{k}_i(J;N) \psi((sd)^{N}_{i};JM) 
\nonumber\\
&+& \
\sum_{j} b^{k}_j(J;N+2)\psi((sd)^{N+2}_{j};JM)~,
\label{eq:wf:U5}
\end{eqnarray}
where $k$, $i$, and $j$ are rank numbers. The  weight of the wave function contained within the 
$[N]$-boson subspace, can then be defined as
the sum of the squared amplitudes $w^k(J,N) \equiv \sum_{i}\mid a^{k}_i(J;N)\mid ^2$. Likewise,
one obtains the content in the $[N+2]$-boson subspace.

\begin{figure}[hbt]
  \centering
  \includegraphics[width=0.5\linewidth]{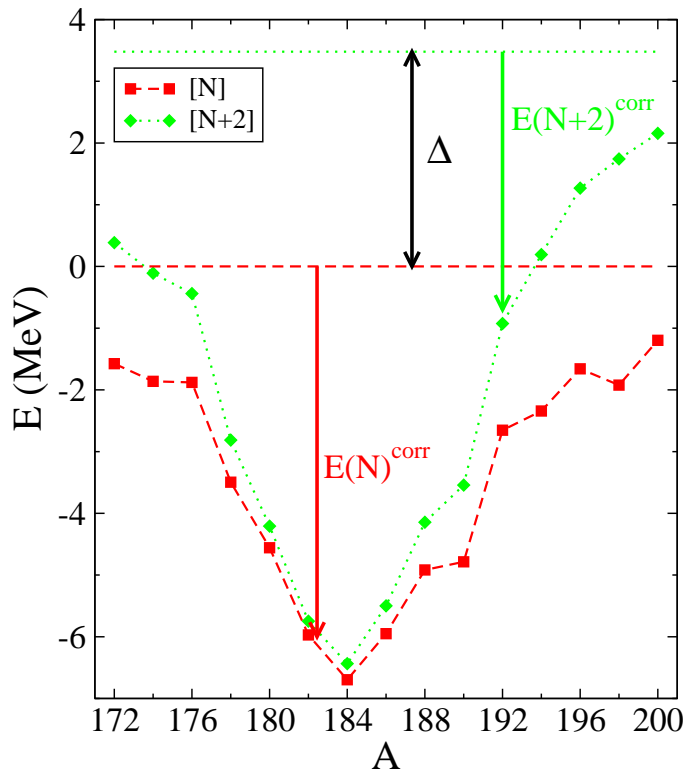}
  \caption{(Color online) Absolute energy of the lowest regular and
    lowest intruder states 
    for $^{172-200}$Hg. The arrows correspond to the correlation
    energies in the N and N+2 subspaces.}
  \label{fig-energ-corr}
\end{figure}

\subsection{The fitting procedure: energy spectra and absolute B(E2) reduced 
transition probabilities}
\label{sec-fit-procedure}

Here, we present the way in which the parameters of the Hamiltonian
(\ref{eq:ibmhamiltonian}), (\ref{eq:cqfhamiltonian}), and (\ref{eq:vmix}) 
and the effective charges in the $\hat{T}(E2)$ transition operator
(\ref{eq:e2operator}) have been determined. 
We study the range $^{172}$Hg to $^{200}$Hg thereby covering
a major part of the neutron N=82-126 shell. 
\begin{table}[hbt]
\caption{Energy levels, characterized by $J^{\pi}_i$, included in
  the energy fit, if known, and the assigned $\sigma$ values in keV.}
\label{tab-energ-fit}
\begin{center}
  \begin{tabular}{|c|c|}
    \hline
    Error (keV) & States  \\
    \hline
    $\sigma=0.1$& $2_1^+$ \\
    $\sigma=1$ & $4_1^+, 0_2^+, 2_2^+$\\
    $\sigma=10$  & $2_3^+, 3_1^+, 4_2^+, 6_1^+, 8_1^+$ \\
    $\sigma=100$   & $2_4^+, 3_1^+, 4_3^+, 6_2^+$\\
    \hline
  \end{tabular}
\end{center}
\end{table}

In the fitting procedure carried out here, we try to obtain the best possible agreement
with the experimental data including both the excitation energies
and the B(E2) reduced transition probabilities. 
Using the expression of the IBM-CM Hamiltonian, as given in equation (\ref{eq:ibmhamiltonian}),
and of the E2 operator, as given in equation (\ref{eq:e2operator}), in the most general case $13$
parameters show up.  
We impose a constraint of obtaining parameters that change smoothly  
in passing from isotope to isotope.  Note also that we constrained 
$\varepsilon_{N+2}=0$, $\kappa'_{N}=0$, and
$\kappa'_{N+2}=0$. We have explored in detail the validity of this
assumption and we have found very little improvement in the value of
$\chi^2$ when releasing those parameters.   
On the other hand, we have kept the value that describes the
energy needed to create an extra particle-hole pair ($2$ extra bosons) constant, 
{\it i.e.}, $\Delta^{N+2}=3480$ keV, and have also put the constraint of keeping the
mixing strengths constant too, {\it i.e.}, $w_0^{N,N+2}=w_2^{N,N+2}=20$ keV for all the Hg isotopes. 
We also have to determine for each isotope the effective charges of the $E2$ operator.
This finally leads to $7$ parameters to be varied in each nucleus. 

To fix the value of $\Delta^{N+2}=3480$ keV we have taken into account
the strong similarity that shows up in Fig.~\ref{fig-energ-hg}
between, on one side (see panel (a)), the Pt spectra resulting from
Refs.~\cite{Garc09,Garc11}, but switching off the mixing term and shifting the 
value of $\Delta$ to  $3480$ keV and, on the other side (see panel
(b)), the experimental energy 
systematics of Hg isotopes. This value of $\Delta^{N+2}$ gives rise
to degenerate $2_1^+$ and $0_2^+$ states at A=182,
which is consistent with the experimental situation observed in the Hg nuclei. To fix the
value of the mixing strengths we considered that the
corresponding value for the Pt nuclei was fixed to $50$ keV \cite{Garc09}, while for
the Pb, to a smaller strength of $18$ keV \cite{fossion03,helle05}. We performed a set
of exploratory calculations between the two latter values and found that the best agreement corresponds
to $w_0^{N,N+2}=w_2^{N,N+2}=20$ keV, although values in the range of $20-30$ keV
provided a very similar agreement.
\begin{table}[hbt]
\caption{Hamiltonian and $\hat{T}(E2)$ parameters resulting from the present study.
 All quantities have the dimension of energy (given in units of keV),
except $\chi_{N+2}$ which is dimensionless and $e_{N}$ and $e_{N+2}$
which are given in units $\sqrt{\mbox{W.u.}}$ 
The remaining parameters of the
Hamiltonian, {\it i.e.}, $\chi_N$, $\varepsilon_{N+2}$, $\kappa'_N$,
and $\kappa'_{N+2}$ are equal to
zero, except  $\Delta^{N+2}=3480$ keV and $w_0^{N,N+2}=w_2^{N,N+2}=20$
keV. }
\label{tab-fit-par-mix}
\begin{center}
\begin{ruledtabular}
\begin{tabular}{cccccccc}
Nucleus&$\varepsilon_N$&$\kappa_N$&$\chi_{N}$&
$\kappa_{N+2}$&$\chi_{N+2}$&$e_{N}$&$e_{N+2}$\\
\hline
$^{172}$Hg& 845.0&   -41.38&  0.01&  -20.70& -1.29&    -&    -\\ 
$^{174}$Hg& 888.6&   -40.21&  0.02&  -19.63&  1.25&    -&    -\\ 
$^{176}$Hg& 906.4&   -34.99&  0.02&  -27.99&  0.01&    -&    -\\ 
$^{178}$Hg&1032.4&   -50.27&  0.15&  -37.56&  0.13&    -&    -\\ 
$^{180}$Hg&1152.1&   -54.39&  0.36&  -38.72& -0.19& 1.38& 2.41\\ 
$^{182}$Hg&1253.4&   -58.46&  0.39&  -39.91& -0.17& 1.11& 2.24\\
$^{184}$Hg&1321.9&   -58.12&  0.41&  -38.74& -0.11& 1.14& 1.94\\
$^{186}$Hg&1097.6&   -56.95&  0.36&  -39.57& -0.16& 1.07& 2.11\\
$^{188}$Hg& 839.4&   -53.17&  0.20&  -38.61& -0.17& 1.42& 2.13\\ 
$^{190}$Hg& 703.3&   -57.59&  0.13&  -42.57&  0.01& 1.42$^*$&2.13$^*$\\
$^{192}$Hg& 697.3&   -42.57&  0.25&  -26.55& -0.60& 1.42$^*$&2.13$^*$\\
$^{194}$Hg& 615.8&   -44.49&  0.19&  -21.34& -1.32& 1.42$^*$&2.13$^*$\\
$^{196}$Hg& 545.9&   -39.79&  0.16&  -18.00& -0.85& 1.81& 2.72$^*$\\
$^{198}$Hg& 449.2&   -54.08&  0.31&  -18.00& -0.85& 1.83& -   \\
$^{200}$Hg& 499.3&   -45.73&  1.07&  -18.00& -0.85& 1.97& -   \\
\end{tabular}
\end{ruledtabular}
\end{center}
\footnotetext{$^*$ The effective charges have been taken the same as
  the corresponding 
  values obtained for $^{188}$Hg, except for $^{196}$Hg where the ratio
  of e$_{N+2}$/e$_N$ was imposed to have
  the same value as in $^{188}$Hg.}
\end{table}

The $\chi^2$ test is used in the fitting procedure in order to extract the optimal solution. 
The $\chi^2$ function is defined in the standard way as
\begin{equation}
  \label{chi2}
  \chi^2=\frac{1}{N_{data}-N_{par}}\sum_{i=1}^{N_{data}}\frac{(X_i
    (data)-X_i (IBM))^2}{\sigma_i^2},
\end{equation} 
where $N_{data}$ is the number of experimental data,
$N_{par}$ is the number of parameters used in the IBM fit, $X_i(data)$
describes the experimental excitation energy of a given excited state (or an experimental 
B(E2) value), $X_i(IBM)$ denotes the corresponding calculated IBM-CM value,
and $\sigma_i$ is an error assigned to each $X_i(data)$ point. 

The $\chi^2$ function is defined as a sum over all data points including excitation
energies as well as absolute B(E2) values. The minimization is carried out using 
$\varepsilon_N$,  
$\kappa_N$, $\kappa_{N+2}$, $\chi_{N}$, $\chi_{N+2}$, $e_{N}$ and
$e_{N+2}$ as free parameters, having fixed
$\varepsilon_{N+2}=0$, $\kappa'_{N}=0$, $\kappa'_{N+2}=0$, $\Delta^{N+2}=3480$
keV and $w_0^{N,N+2}=w_2^{N,N+2}=20$ keV as described before. We minimize the $\chi^2$
function for each isotope separately using the package MINUIT \cite{minuit} which allows to
minimize any multi-variable function.
In some of the lighter Hg isotopes, due to the small number of experimental data, the
values of some of the free 
parameters could not be fixed unambiguously using the above fitting
procedure. Moreover, for the heavier isotopes ($A>194$), that part of the
Hamiltonian corresponding to the intruder states is fixed such as to guarantee 
that those states appear well above the regular ones, that
is, above $3$ MeV. In some cases due to the lack of experimental data
the effective charges could not be determined. For $A>196$, $e_{N+2}$
cannot be determined because the B(E2) values are insensitive to this
parameter. However, for completeness we have taken the effective
charges of $^{190-194}$Hg equal to the ones of $^{188}$Hg while $e_{N+2}$ in
$^{196}$Hg was obtained by imposing the constraint to have the same ratio
$e_{N+2}/e_{N}$ as for $^{188}$Hg. 
\begin{figure}[hbt]
  \centering
  \includegraphics[width=0.8\linewidth]{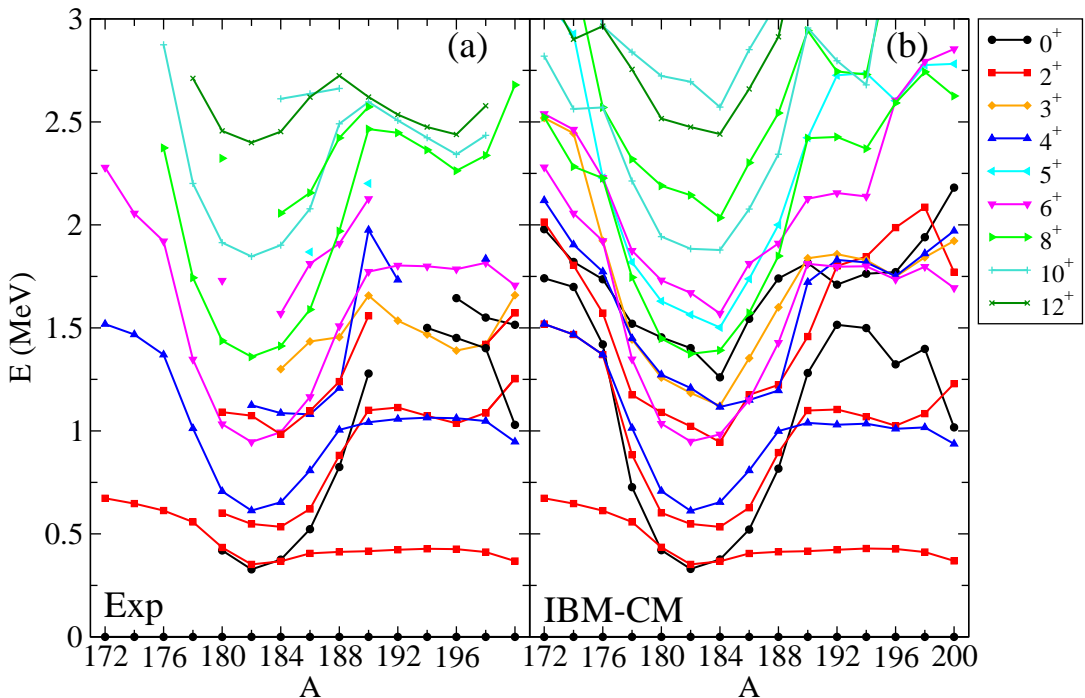}
  \caption{(Color online) Experimental excitation energies (up to $E_x \approx 3.0$
    MeV) (panel (a)) and the theoretical results (panel (b)),
    obtained from the IBM-CM.} 
  \label{fig-energ-comp}
\end{figure}

As input values, we have used the excitation energies of the levels presented in 
table \ref{tab-energ-fit}. In this table we also give the corresponding $\sigma$ values. 
We stress
that the $\sigma$ values do not correspond
to experimental error bars, but they are related with the expected
accuracy of the IBM-CM calculation to reproduce a particular experimental data point. 
Thus, they act as a guide so that a given calculated level converges towards the
corresponding experimental level.
The $\sigma$ ($0.1$ keV) value for the $2_1^+$ state guarantees
the exact reproduction of this experimental most important excitation energy, {\it
  i.e.}, the whole energy spectrum is normalized to this experimental
energy. The states $4_1^+, 0_2^+$ and $2_2^+$ are 
considered as the most important ones to be reproduced ($\sigma=1$
keV). The group of states  
$2_3^+$, $3_1^+$, $4_2^+$, $6_1^+$, and $8_1^+$ ($\sigma=10$
keV) and  $2_4^+$, $3_1^+$, $4_3^+$, and $6_2^+$ ($\sigma=100$ keV) should also be well
reproduced by the calculation to guarantee a correct moment of
inertia for the yrast band and the structure of the pseudo-$\gamma$
and $0_2^+$ bands. Besides, in specific nuclei, A $=186, 190,$ and 196, additional states
have been taken into account, but in all cases
those states have an excitation energy below $2-2.5$ MeV and $J<10$. Note also
that if two, or more, angular momenta are assigned to a given level (see the references 
~\cite{Kibedi10,Browne99,Basu06,Achter09,Wu03,Sing11,Bagl10,Bagl03,Sing02,Sing03,Bagl12,Sing06,Xia07,Xia02,Kondev07} 
and/or the extra references given in Section ~\ref{sec-expe}),
those levels are not included in the fitting  procedure.

In the case of the $E2$ transitions rates, we have used the available
experimental data involving the states presented in table \ref{tab-energ-fit},
restricted to those $E2$ transitions for which absolute B(E2) values are known.
Additionally we have taken a value of $\sigma$ that corresponds to $10\%$ of the 
B(E2) values or to the experimental error bar if larger, except
  for the transition $2_1^+\rightarrow 0_1^+$ where a smaller value of
  $\sigma$ ($0.1$ W.u.) was taken, therefore normalizing to the
  experimental $B(E2;2^+_1 \rightarrow 0^+_1)$ value. 
The experimental data we have used result from the
data appearing in references ~\cite{Kibedi10,Browne99,Basu06,Achter09,Wu03,Sing11,Bagl10,Bagl03,Sing02,
Sing03,Bagl12,Sing06,Xia07,Xia02,Kondev07},
complemented with the specific references already presented in Section \ref{sec-expe}.
In the present fit, we have not included relative B(E2)
values, which may well slightly modify the effective charges obtained
at present. 

This has resulted in the values of the parameters for the IBM-CM Hamiltonian,
as given in table \ref{tab-fit-par-mix}. 
In the case of $^{172-178}$Hg and $^{190-196}$Hg, 
the value of the effective charges, or part of them, cannot be determined
because not a single absolute B(E2) value is known or $\chi^2$ is
insensitive to their values. However, for
completeness we have taken the value of the effective charges in  $^{190-194}$Hg.
to be the same as in $^{188}$Hg (or as having the same ratio for $^{196}$Hg).
\begin{figure}[hbt]
  \centering
  \includegraphics[width=0.6\linewidth]{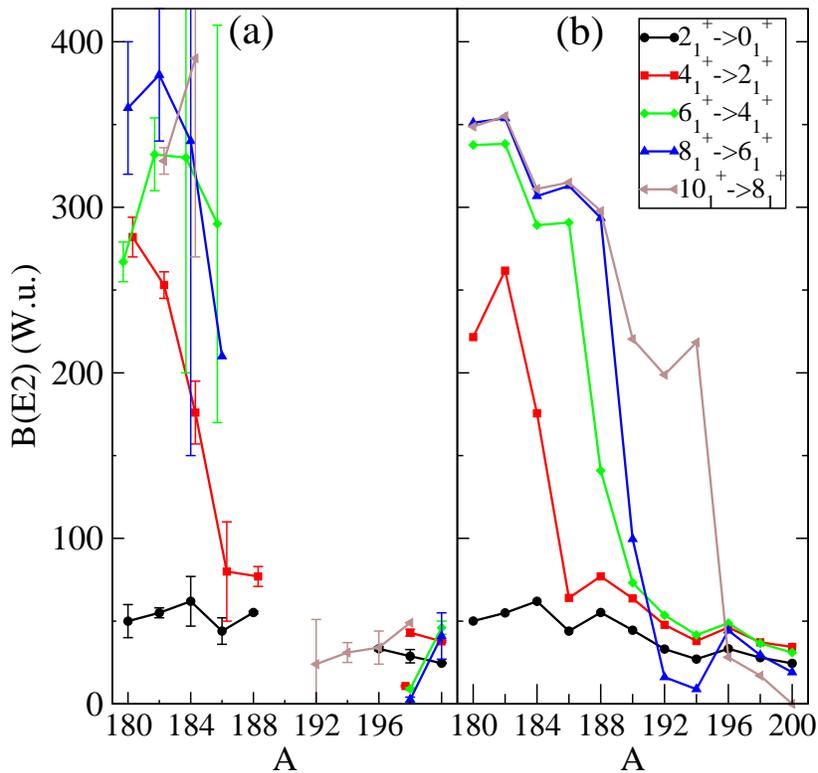}
  \caption{(Color online) Comparison of the absolute B(E2) reduced transition
    probabilities along the yrast band, given in W.u. The panel (a) corresponds to known
    experimental data and the panel (b) to the theoretical IBM-CM
    results.} 
  \label{fig-be2-1}
\end{figure}

\subsection{Correlation energy in the configuration mixing approach} 
\label{sec-corr_energy}

Intruder configurations should appear, by construction, at an excitation energy that is 
much higher than the regular configurations. This is so because of the large energy
needed to create the 2p-2h excitation across the Z=82 closed shell. In the case of the Hg nuclei,
$\Delta^{N+2}=3480$ keV, but according to reference \cite{hey87} the single-particle energy cost
has to be corrected because of the strong pairing energy gain when forming two extra $0^+$ coupled (particle and hole)
pairs, the quadrupole energy gain when opening up the proton shell, as well as 
by the monopole correction caused by a change in the single-particle energy gap at Z=82 as a function of the neutron
number. In some cases, specifically around the mid-shell point at N=104, the energy gain through these
correlations can become so large that the intruder configurations intrude to be located below the
energy of the regular configurations. In this case one speaks about ``islands of inversion'' \cite{heyde11}.

A different way to understand the relative position of regular and
intruder configurations is to consider the energy of the lowest lying
regular and intruder state. The regular configuration,
which corresponds to a spherical or slightly deformed shape, can be
considered as the ``reference'' state and to have zero energy. This configuration will be lowered, as a function of
neutron number, because of the correlation energy due to
the quadrupole interaction (in our case using the IBM). On the other
hand, the intruder configuration corresponds to a more strongly deformed shape.
Its energy will then be equal to $\Delta^{N+2}$ corrected by the
correlation energy, this time within the (N+2) configuration. This situation is illustrated in
Fig.~\ref{fig-energ-corr} where it is clearly appreciated how the
energies of both configurations can come very close in energy, depending on the balance between the
off-set $\Delta^{N+2}$ and the difference in the correlation energy $E(N+2)^{corr} - E(N)^{corr}$.

One observes that around mid shell, both configurations are fairly
close in energy although the energy of the regular configuration is
below the intruder configuration in all cases. 
However, near the beginning and the end of the shell, this energy difference becomes much larger. 
Note that the value of $\Delta^{N+2}$ constrains the parabolic shape of the intruder energy systematics 
at both sides of the shell. Near the doubly-closed shells, the $0^+$ ground state is approximately 
spherical and the corresponding correlation energy will be small. 
Therefore, the largest possible excitation energy for the lowest intruder configuration should appear at
an energy below $\Delta^{N+2}$, simply  because the correlation energy for the intruder configuration
(having two more nucleon pairs) is larger than the correlation energy for the regular states. 
Consequently, the parabolic behavior which shows up around mid shell will
be eroded for the lightest and the heaviest isotopes, 
resulting in a rather flat energy systematics and therefore intruder
configurations result at an energy lower than expected.
\begin{figure}[hbt]
  \centering
  \includegraphics[width=0.6\linewidth]{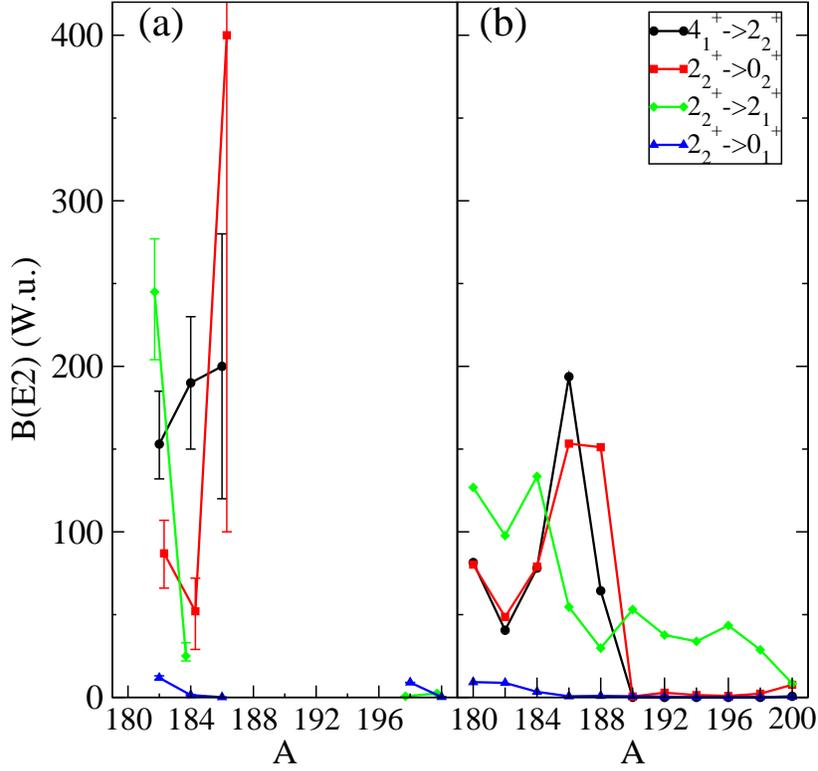}
  \caption{(Color online) Comparison of the few non-yrast absolute B(E2) reduced transition
    probabilities, given in W.u. The panel (a) corresponds to the few known
    experimental data, the panel (b) to the theoretical IBM-CM results.}
  \label{fig-be2-2}
\end{figure}

\subsection{Detailed comparison between the experimental data and the IBM-CM results:
 energy spectra and electric quadrupole properties}
\label{sec-fit-compare}

In the present subsection, we compare the 
experimental energy spectra with the energy spectra as obtained from
the IBM-CM, 
for the limited data set (with excitation energy $E_x$ less than
$\approx$ 3.0 MeV), so as to be able to carry out a detailed
comparison of the experimental data (panel (a) of Fig.~\ref{fig-energ-comp}) and the calculations
(panel (b) of Fig.~\ref{fig-energ-comp}). 
In comparing both panels of Fig.~\ref{fig-energ-comp} one can observe
a rather good overall agreement. Note that the energy of the $2_1^+$
level matches perfectly the experimental one because we used precisely
this level to constrain the calculation. For the other levels we
reproduce correctly the observed almost parabolic behavior of the energy levels
with a lowest excitation energy of the second $0^+_2$ state at N=102 (A=182) (near neutron mid-shell), while a rather 
flat behavior of the excitation energies as a function of increasing neutron number for the heavier Hg nuclei
(A $\geq$ 190) shows up, consistent with the data. For the nuclei
with mass number below A=180 (172 $\leq$ A $\leq$ 180), we perfectly match
the steady increase of the energy levels with decreasing mass number A.
We point out that for the nuclei near mid-shell, the experimental energy systematics is
reproduced up to angular momentum $J=12$, although states with $J=10$ and $J=12$ have not been
included in the fitting procedure.
In general, the agreement with the data is better for the states with even $J$ values,
while the calculated theoretical energy levels with odd $J$ values appear systematically above the
experimental excitation energies.

We carry out a more detailed comparison for both the energies and E2 properties in the region where shape
coexistence shows up most clearly (180 $\leq$ A $\leq$ 188) in the later part of this section.

A more stringent test than a good reproduction of the energy
systematics comes from calculating those observables that probe the corresponding wave functions
and a comparison with the data. 
Recent experimental efforts have given rise to absolute B(E2) values along the yrast band, 
deduced from lifetime measurements~\cite{grahn09,scheck10}, as well as
through Coulomb 
excitation at ISOLDE-CERN~\cite{Isol12,lipska13},
the latter also giving first results for the quadrupole moments of the $2^+_1$ and $2^+_2$ states.

\begin{figure}[hbt]
  \centering
  \includegraphics[width=0.5\linewidth]{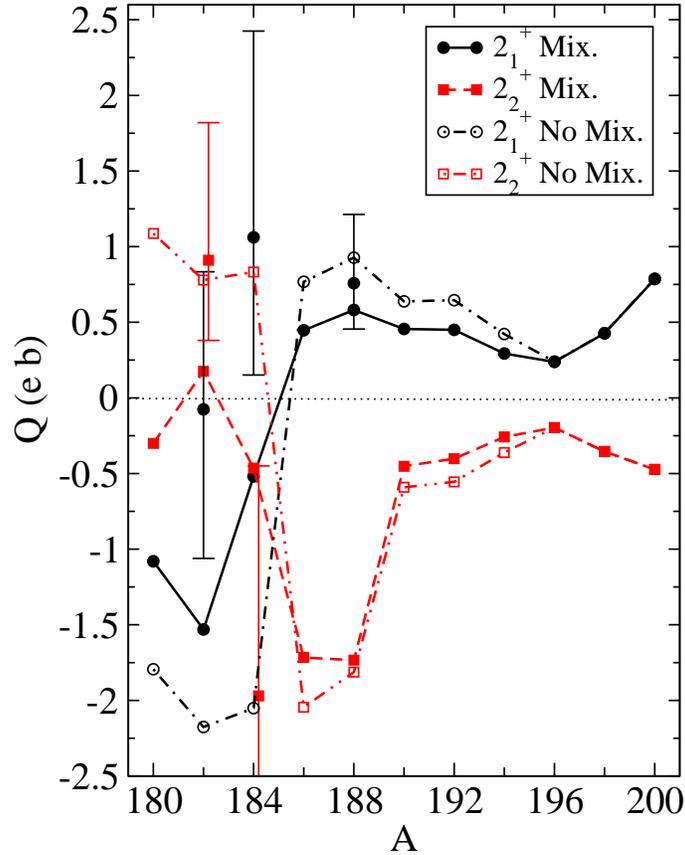}
  \caption{(Color online) IBM-CM values of the quadrupole moments of the states
    $2_1^+$ and $2_2^+$ for $^{180-200}$Hg. Quadrupole moments given
    in units $e\, b$. The dash-dotted lines indicate the quadrupole moments when the mixing Hamiltonian is
    switched off, corresponding to the unperturbed states.}
  \label{fig-q-theo}
\end{figure}

The systematics for a number of important E2 transitions and the electric quadrupole
moments are presented in figures
\ref{fig-be2-1}, \ref{fig-be2-2}, and \ref{fig-q-theo}, respectively.

Because these figures highlight only the specific set of important B(E2) values, we detail
the comparison between the present theoretical study and the existing
data (which is most 
documented in the 180 $\leq$ A $\leq$ 190 region of shape coexistence)
in tables \ref{tab-be2a}, \ref{tab-be2b}, and \ref{tab-be2-1}
in which we compare the experimental absolute B(E2) values as well as the relative B(E2) values, respectively, with the 
corresponding IBM-CM theoretical results. 

In Fig.~\ref{fig-be2-1} (panel (a)) one observes a rather constant
value for the $B(E2;2^+_1 \rightarrow 0^+_1)$ at approximately
$\approx$ 50 W.u., 
and very large B(E2) values for the higher spin values, with a particular interesting behavior for the
$B(E2;4^+_1 \rightarrow 2^+_1)$ which is dropping from A=182 towards a more stable value (see
also the calculated variation in Fig.~\ref{fig-be2-1} (panel (b))) starting at A=186 and onwards. This very clearly shows an important change in the structure
of the $2^+_1$ and $2^+_2$ states passing through the region A=180 to A=188. We come back later to this most
important observation, that highlights a particular mixing pattern between those $2^+$ states.  
Concerning the high-spin $10^+_1$ to $8^+_1$ E2 transition in 
the region A=190 to A=194, a pronounced collective character still exits.
This is indeed what is expected in the IBM. In the case of $^{196-198}$Hg the agreement is improved due to   
the reduction in the number of available bosons. 

In Fig.~\ref{fig-be2-2} (panel (a)), we show the few non-yrast absolute B(E2)
values known (see also table \ref{tab-be2-1}). Here, one observes
larger $B(E2;4^+_1 \rightarrow 2^+_2)$ values as compared to the
corresponding yrast B(E2) value (starting at A=186) which is again a
clear hint of the changing mixing between the unperturbed
configurations making up the lowest two $2^+$ states.  The comparison
with the theoretical values (see Fig.~\ref{fig-be2-2} panel (b)),
where a maximal $B(E2;4^+_1 \rightarrow 
2^+_2)$ value is obtained for A=184 looks interesting, the more
because the $B(E2;2^+_2 \rightarrow 0^+_2)$ exhibits a similar
behavior, pointing out a very specific change in the composition of
the $2^+$ states (see also Section ~\ref{sec-evolution} for a more extensive
discussion).

The above observations are most interesting because one observes a
smooth behavior in the excitation energy of the $2^+_{1,2}$ states
when moving from A=180 towards A=188, still there must be a major
change in the wave functions of these states.  The particular mixing
between the $2^+_{1,2}$ regular and intruder configuration is
dramatically present in the calculated spectroscopic quadrupole
moments as shown in Fig.~\ref{fig-q-theo}. Up to mass A=184, the
nucleus in which our calculations result in a close to equal mixing
between the regular and intruder configurations (see
Fig.~\ref{fig-wf}), the intruder configuration dominates in the
$2^+_1$ state, giving rise to the negative sign, and opposite to the
$2^+_2$ state. For masses above A=184, the calculations result in a
$2^+_1$ state with increasing weight for the regular configuration,
and the opposite for the $2^+_2$ state, up to A=188. From mass A=190
onwards, both $2^+_{1,2}$ are described by wave functions within the
regular configuration space only. We complement this figure by also including
the quadrupole moments for the unperturbed lowest intruder and regular
state (obtained from switching off the mixing Hamiltonian).

\begin{figure}[hbt]
  \centering
  \includegraphics[width=1\linewidth]{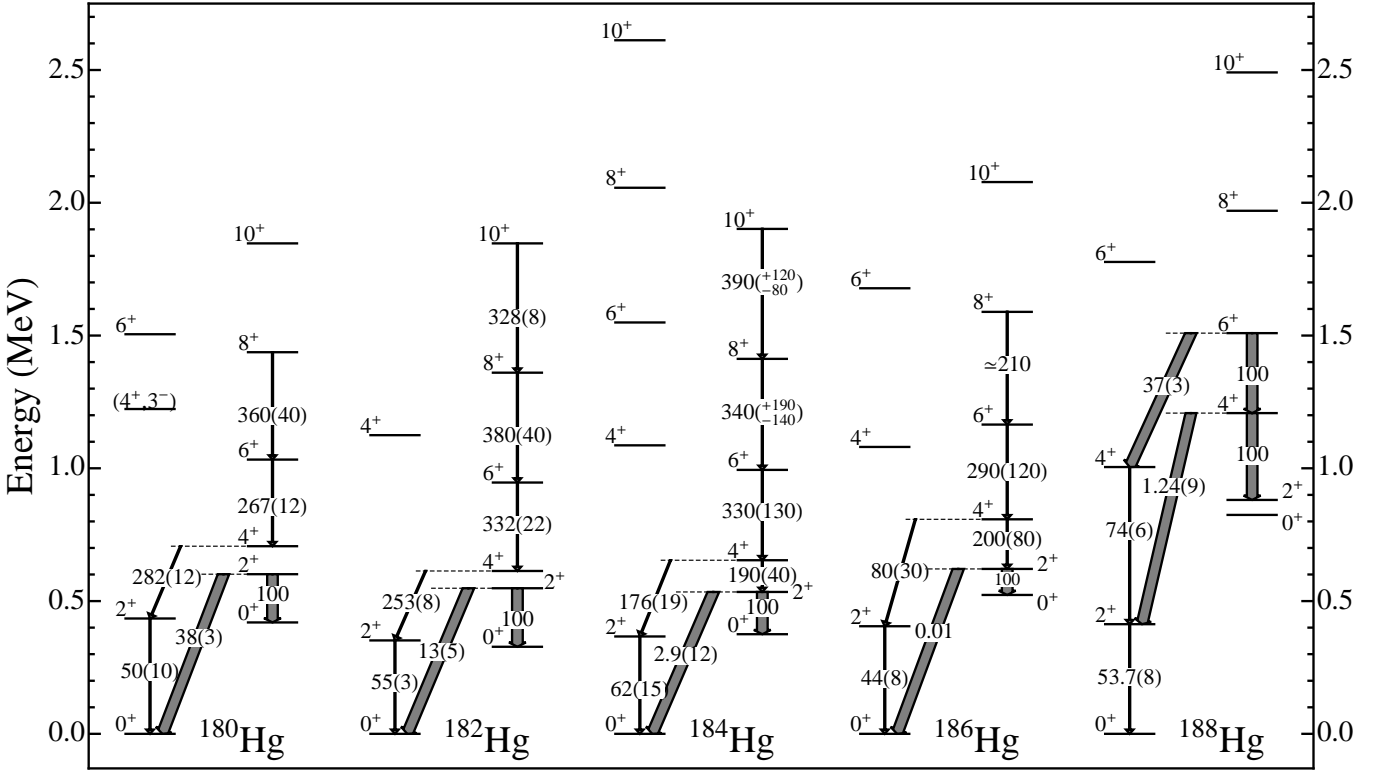}
  \caption{Experimental excitation energies, absolute B(E2)
    transition rates and relative B(E2) values for selected
    states in $^{180-188}$Hg. Thin lines correspond to absolute
    B(E2) values while the thicker ones to relative B(E2) values.} 
  \label{fig-exp-180-188}
\end{figure}

\begin{figure}[hbt]
  \centering
  \includegraphics[width=1\linewidth]{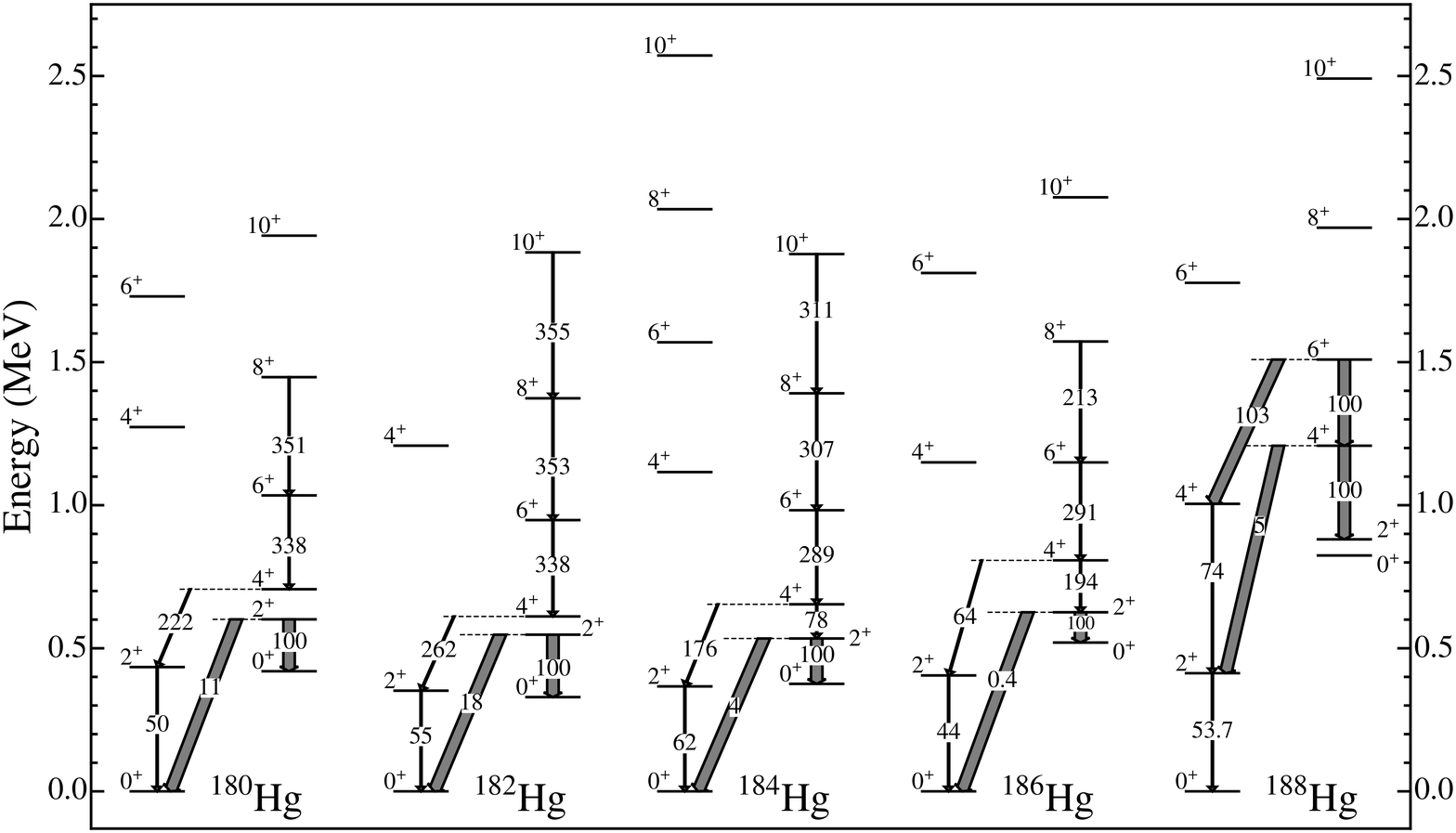}
  \caption{Theoretical excitation energies, absolute B(E2)
    transition rates and relative B(E2) values for selected
    states in $^{180-188}$Hg. Thin lines correspond to absolute
    B(E2) values while the thicker ones to relative B(E2) values.} 
  \label{fig-theo-180-188}
\end{figure}

We present in figures \ref{fig-exp-180-188} and \ref{fig-theo-180-188} 
the experimental and theoretical energy spectra, respectively (up to $E_x=2-3$ MeV), concentrating on the region
where the ``coexistence'' of two structures, with their specific energy
scale, becomes obvious: one less and one more deformed configuration.
In Fig.~\ref{fig-exp-180-188}, we combine the
energies and the experimental B(E2) values. We give the absolute B(E2) values when
known and the relative ones, mainly at the low spin part of the energy spectra.

A first point is the fact that the proximity of the $2^+_1$ and $2^+_2$
states within an energy of $\approx$ 180 keV from A=180 up to A=184
indicates rather strong mixing between the two configurations. An indicator
for the mixing shows up from the relative B(E2) values de-exciting the $2^+_2$
state. Using a normalization to 100 for the $B(E2; 2^+_2 \rightarrow 0^+_2)$,
the value of $B(E2; 2^+_2 \rightarrow 0^+_1)$ 
 decreases from $38(3)$ W.u.~(A=180), $13(5)$ W.u.~(A=182), $2.9(1.2)$ W.u.~(A=184)
down to $0.01$ W.u.~(A=186), moving quickly down for heavier masses.
The observation of a strong E0 component in the decay of the $2^+_2$
state into the $2^+_1$ state ~\cite{page11,scheck11} 
is a quantitative indicator of important mixing in the wave functions ~\cite{joshi94,wood99}.
This important information, extended with the large absolute B(E2) values
originating from the yrast $10^+$,...,$6^+$ states, allows us to separate the
levels into two families. This separation is substantiated by the much smaller and
fairly constant $B(E2;2^+_1 \rightarrow 0^+_1)$ value $\approx$ 50 W.u.

We also notice that the E2 decay from the $4^+_1$ into the $2^+_{1,2}$ levels
is such that from mass A=184 onwards, the $B(E2;4^+_1 \rightarrow
2^+_2)$ value (which is still almost equal to the  
$B(E2;4^+_1 \rightarrow 2^+_1)$ value) starts to dominate quickly when moving
towards mass A=188, with relative B(E2) values changing from 200(80)
over 80(30)  for A=186, 
up to $100$ over $1.24(9)$ for A=188. This indicates that the
intruder structure is quickly moving up in energy 
relative to the regular structure.  
Combining all the available data: excitation energies, absolute and
relative B(E2) values, it shows that only in A=186 
the coupling between, on one side the $0^+_1, 2^+_1, 4^+_2$ states, and, on the
other side the $0^+_2,2^+_2,4^+_1$ states, as expressed by the weights
$w^k(J,N)$ (see expression (\ref{eq:wf:U5})), comes out to be less
than $20\%$ (see Fig.~\ref{fig-wf}) consistent with rather weakly coupled bands.  
 
In A=188, perturbations are arising for the $4^+$ and $6^+$ levels,
as indicated by the relative B(E2) values originating from the decay of the $6^+_1$ state, having
a clear preference for the $4^+_2$ level. Here it looks like the $4^+_1$ state has
become a member of the less collective band. 

In view of the above analyses of the experimental available data, the present
separation into two families can be made. However, the relative changes in energy
differences at the lower end of the intruder band as well as the relative B(E2)
values, connecting the two bands, unambiguously show important mixing between the
$2^+$ members of the two families.

In Fig.~\ref{fig-theo-180-188}, we compare with the corresponding theoretical energy spectra and B(E2) values,
denoting the absolute values and relative B(E2) values, in order to allow an
easy comparison with the data shown in Fig.~\ref{fig-exp-180-188}.
The overall structure, both on energy spectra as well as on the absolute B(E2) values, agrees rather
well with the corresponding experimental figure. Because at the lower end of the intruder
band, the deexcitation largely favors decay into the $2^+_1$ state rather than into the $2^+_2$ state
and thus makes it very difficult to obtain absolute B(E2) values, we show the relative B(E2) values,
normalized at 100 for the  $B(E2;2^+_2 \rightarrow 0^+_2)$ value. One observes a steady increase of
the ratio from A=188 down to A=180, consistent with the experimental data, with a clear dominance
of the inband E2 decay. This is consistent with the fact that the wavefunction of the ground state
$0^+_1$ state is mainly of regular character, the $0^+_2$ mainly of intruder character,
whereas the nature of the $2^+_2$ exhibits a global increase of its regular character going from
A=188 down to A=182 (see Fig.~\ref{fig-wf}).      

\begin{figure}[hbt]
  \centering
  \includegraphics[width=0.5\linewidth]{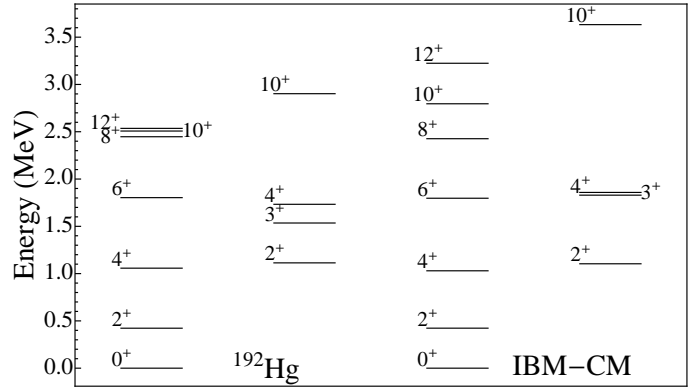}
  \caption{ Detailed comparison between the experimental energy spectrum 
and the calculated IBM-CM spectrum for $^{192}$Hg.} 
  \label{fig-192hg}
\end{figure}

Moving into the heavier Hg isotopes (beyond A=190), 
showing the experimental and theoretical spectra of  $^{192}$Hg (Fig.~\ref{fig-192hg}) as an example, 
the occurrence of shape coexistence seems to be dissolved.

Inspecting the systematics of the Hg nuclei, from A=188 and onwards up
to A$\approx 200$, the
appearance of a set of close-lying high-spin states, \textit{i.e.}, the $12^+,10^+$ and $8^+$
state is striking (see Fig.~\ref{fig-system-hg}). 
From measurements of the magnetic moment of the $12^+_1,10^+_1$ states
in the nuclei A=188 to A=196 ~\cite{Stone05}, the deduced g-factor
varies between $-0.19(11) \mu_N$
and $-0.24(4) \mu_N$, which corresponds very well with the g-factor characterizing the neutron $\nu 1i_{13/2}$
single-particle orbital, pointing towards a clear-cut one broken-pair character of $(1i_{13/2})^2;J^{\pi}$ structure.
The slight rise in excitation energy approaching the neutron N=126 shell closure is consistent
with the picture of a state becoming less collective and more shell-model like in its character.  

\begin{figure}[hbt]
  \centering
  \includegraphics[width=0.4\linewidth]{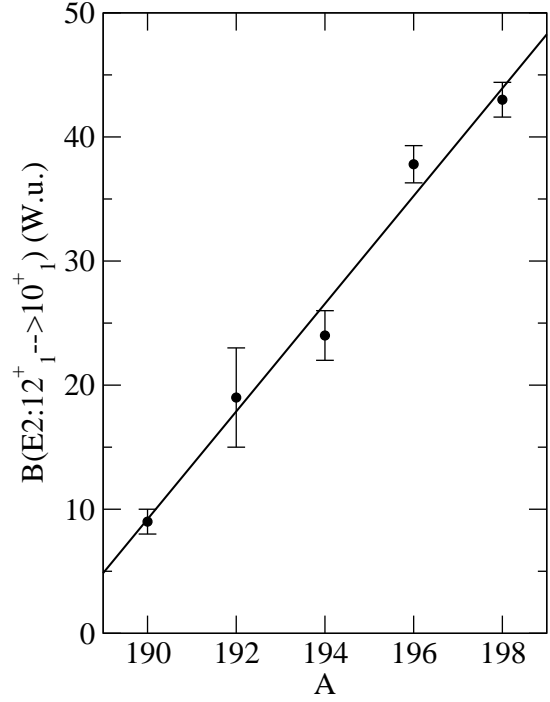}
  \caption{Experimental values of $B(E2;12^+_1 \rightarrow 10^+_1)$. Data are taken from the corresponding
  NDS references \cite{Sing03,Bagl12,Sing06,Xia07,Xia02}.} 
  \label{fig-be2-12-10}
\end{figure}

A na\"{i}ve shell-model point-of-view indeed shows the filling of the $\nu 1i_{13/2}$ orbital in between
$^{180}$Hg and $^{194}$Hg. Because of pairing correlations, one can expect an effect on the E2 transition
probability connecting the $12^+_1$ state with the $10^+_1$ state. Experimental data are available on this
transition, giving rise to B(E2) values and partly for the $10^+_1$ state decaying to the $8^+_1$ state
(see A=190 \cite{Sing03}, A=192 \cite{Bagl12}, A=194 \cite{Sing06}, A=196 \cite{Xia07}, A=198 \cite{Xia02}).
For the $B(E2;12^+_1 \rightarrow 10^+_1)$ reduced transition probability, one notices a decreasing value
from the largest known value of 43(12) W.u.~(in A=198) coming down to 9(1) W.u.~(in A=190). This variation, as
shown in Fig.~\ref{fig-be2-12-10}, looks
quite linear with A and might reflect the effect of the occupation of the $1i_{13/2}$ neutron orbital, which,
for a pure seniority non-changing $v=2 \rightarrow v=2$ E2 transition would change with the factor
$[u^2(1i_{13/2}) - v^2(1i_{13/2})]$. 

\begin{table}
  \caption{Comparison of the experimental absolute B(E2) values (given in
    units of W.u.) with
    the IBM-CM Hamiltonian results.
    Data are taken from  the Nuclear Data
    Sheets~\cite{Kibedi10,Browne99,Basu06,Achter09,Wu03,Sing11,Bagl10,Bagl03, 
      Sing02,Sing03,Bagl12,Sing06,Xia07,Xia02,Kondev07},  
    complemented with references presented in section
    \ref{sec-expe}.}  
  \label{tab-be2a}
\begin{center}
\begin{ruledtabular}
\begin{tabular}{cccc}
Isotope   &Transition             &Experiment&IBM-CM \\
\hline
$^{180}$Hg&$2_1^+\rightarrow 0_1^+$& 50(10)         & 50    \\  
         &$4_1^+\rightarrow 2_1^+$& 282(12)       & 223 \\   
         &$6_1^+\rightarrow 4_1^+$& 267(12)       & 338   \\   
         &$8_1^+\rightarrow 6_1^+$& 360(40)       & 351   \\   
\hline
$^{182}$Hg&$2_1^+\rightarrow 0_1^+$& 55(3)         & 55    \\  
         &$4_1^+\rightarrow 2_1^+$& 253(8)       & 262 \\   
         &$6_1^+\rightarrow 4_1^+$& 332(22)       & 338   \\   
         &$8_1^+\rightarrow 6_1^+$& 380(40)       & 354   \\   
         &$10_1^+\rightarrow 8_1^+$& 328(8)*      & 355   \\   
         &$2_2^+\rightarrow 0_1^+$& 11.8(12)*\footnotemark[1]      &  9  \\   
         &$2_2^+\rightarrow 0_2^+$& 87$(^{+20}_{-23})$*\footnotemark[1]      &  49  \\   
         &$2_2^+\rightarrow 2_1^+$& 245$(^{+32}_{-41})$*\footnotemark[1]      & 98   \\   
         &$4_1^+\rightarrow 2_2^+$& 153$(^{+32}_{-21})$*\footnotemark[1]        &  41  \\   

\hline 
$^{184}$Hg&$2_1^+\rightarrow 0_1^+$&  62(15)     &  62 \\  
         &$4_1^+\rightarrow 2_1^+$&  176(19)    &  176\\   
         &$6_1^+\rightarrow 4_1^+$&  330(130)   &  289  \\   
         &$8_1^+\rightarrow 6_1^+$&  340$(^{+190}_{-140})$    &  306  \\   
         &$10_1^+\rightarrow 8_1^+$& 390$(^{+120}_{-80})$*    &  311  \\   
         &$2_2^+\rightarrow 0_1^+$& 1.4(3)*\footnotemark[1]      &   3.4 \\   
         &$2_2^+\rightarrow 0_2^+$& 52(20)*\footnotemark[1]      &  79  \\   
         &$2_2^+\rightarrow 2_1^+$& 25(8)*\footnotemark[1]      &  133  \\   
         &$4_1^+\rightarrow 2_2^+$&  190(40)  and  500(80)*\footnotemark[1] &  78\\   
\hline 
$^{186}$Hg&$2_1^+\rightarrow 0_1^+$&  44(8)     &  44 \\  
         &$4_1^+\rightarrow 2_1^+$&  80(30)    &  64\\   
         &$6_1^+\rightarrow 4_1^+$&  290(120)   &  291  \\   
         &$8_1^+\rightarrow 6_1^+$&  $\approx 210$    &  313  \\   
         &$2_2^+\rightarrow 0_1^+$& $(0.0,+0.2)$*\footnotemark[1]      &  0.6  \\   
         &$2_2^+\rightarrow 0_2^+$&  400(300) and $>140$*\footnotemark[1]  &  153\\   
         &$4_1^+\rightarrow 2_2^+$&  200(80)  and  490$(^{+240}_{-100})$*\footnotemark[1] &  194\\  
\end{tabular}
\end{ruledtabular}
\end{center}
\footnotetext{$^*$ Experimental data not included in the fit.}
\footnotetext[1] {Data taken from \cite{lipska13}.}
\end{table}

\begin{table}
  \caption{See caption of Table \ref{tab-be2a}.}  
  \label{tab-be2b}
\begin{center}
\begin{ruledtabular}
\begin{tabular}{cccc}
Isotope   &Transition             &Experiment&IBM-CM \\

\hline 
$^{188}$Hg&$2_1^+\rightarrow 0_1^+$&  53.7(8)     &  54 \\  
         &$4_1^+\rightarrow 2_1^+$&  74(6)    &  74\\   
\hline 
$^{190}$Hg&$(12_1^+)\rightarrow (10_1^+)$&  9(1)*    &  216**\\   
\hline 
$^{192}$Hg&$(10_1^+)\rightarrow 8_1^+$& 24$(^{+27}_{-24})$* &  199**\\ 
        &$(12_1^+)\rightarrow 10_1^+$& 19(4)* &  186**\\   
\hline 
$^{194}$Hg&$(10_1^+)\rightarrow 8_1^+$& 31(6)* &  218**\\   
        &$(12_1^+)\rightarrow (10_1^+)$& 24(2)* &  193**\\   
\hline 
$^{196}$Hg&$2_1^+\rightarrow 0_1^+$&  33.3(12)     &  33.3 \\  
         &$(10_1^+)\rightarrow (8_1^+)$&  34(10)    &  33\\   
         &$(12_1^+)\rightarrow (10_1^+)$&  37.8(15)*    &  35**\\   
\hline 
$^{198}$Hg&$2_1^+\rightarrow 0_1^+$&  28.8(4)     &  27.9 \\  
         &$4_1^+\rightarrow 2_1^+$&  43(2) and 10.8(5) * &  37\\   
         &$6_1^+\rightarrow 4_1^+$&  9.0(8)     &  37  \\   
         &$8_1^+\rightarrow 6_1^+$&  2.6(15)   &  30  \\   
         &$10_1^+\rightarrow 8_1^+$&  $\approx 49$* &  17\\   
         &$12_1^+\rightarrow 10_1^+$&  43.0(14)* &  17**\\   
         &$2_2^+\rightarrow 2_1^+$&  0.63(8)  &  29\\   
         &$2_2^+\rightarrow 0_1^+$&  0.0217(5)*  &  0.32\\   
\hline 
$^{200}$Hg&$2_1^+\rightarrow 0_1^+$&  24.57(22)     &  24.5 \\  
         &$4_1^+\rightarrow 2_1^+$&  37.8(6)   &  34\\   
         &$6_1^+\rightarrow 4_1^+$&  46(4)     &  31  \\   
         &$8_1^+\rightarrow 6_1^+$&  41(14)   &  19  \\   
         &$2_2^+\rightarrow 2_1^+$&  2.4(6)  &  8.4\\   
         &$2_2^+\rightarrow 0_1^+$&  0.23(6)  &  0.36\\   
\end{tabular}
\end{ruledtabular}
\end{center}
\footnotetext{$^*$ Experimental data not included in the fit.}
\footnotetext{$^{**}$ The effective charges have been taken the same as
  the corresponding 
  values obtained for $^{188}$Hg (see also table \ref{tab-fit-par-mix}). }
\end{table}

High-spin states have been incorporated in an extension of the IBM which allows for one and two broken pairs
thereby combining both the collective and specific few-nucleon effects ~\cite{Iac91}. This extension
results in a wave functions of the form $\mid N \rangle \oplus \mid N-1,2qp \rangle \oplus \mid N-2,4qp \rangle$ 
incorporating both collective as well as the important 2qp configurations when applied to the Hg nuclei.
Calculations, covering the A=190-194 Hg nuclei were carried out by
Vretenar {\it et al.}~\cite{Vret93},  
concentrating on the energy spectra and,
more in particular, on the E2 transitions between the $12^+_1$ and $10^+_1$ as well as between the $10^+_1$ and
$8^+_1$ state. It becomes clear though that the too low theoretical values point to an
underestimation of the collective component definitely contributing to the E2 transition. As a reference, 
calculating $B(E2;12^+_1 \rightarrow 10^+_1)$ for a pure $(1i_{13/2})^2$ configuration, a value of
22 $e^2\cdot fm^4$ (2.7 W.u.) results (using an effective neutron charge of
1 e). Similar high-spin structures accentuating
the even higher-spin backbending using the IBM have been performed for
the whole region A=184-200 \cite{Hsieh92}.

\subsection{The evolution of the character of the yrast band: simple configurations versus configuration-mixing}
\label{sec-evolution}

Even though the presence of the shape coexisting structures looks compelling, we analyze in the present section
in quite some detail, how the wave functions describing the coupling amongst the two sets of underlying configurations
is changing going through the long series of the Hg isotopes.

We start our analysis with the structure of the configuration-mixed wave functions 
along the yrast levels, expressed 
as a function of the $[N]$ and $[N+2]$ basis states, as given in eq.~(\ref{eq:wf:U5}). 
\begin{figure}[hbt]
  \centering
  \includegraphics[width=.6\linewidth]{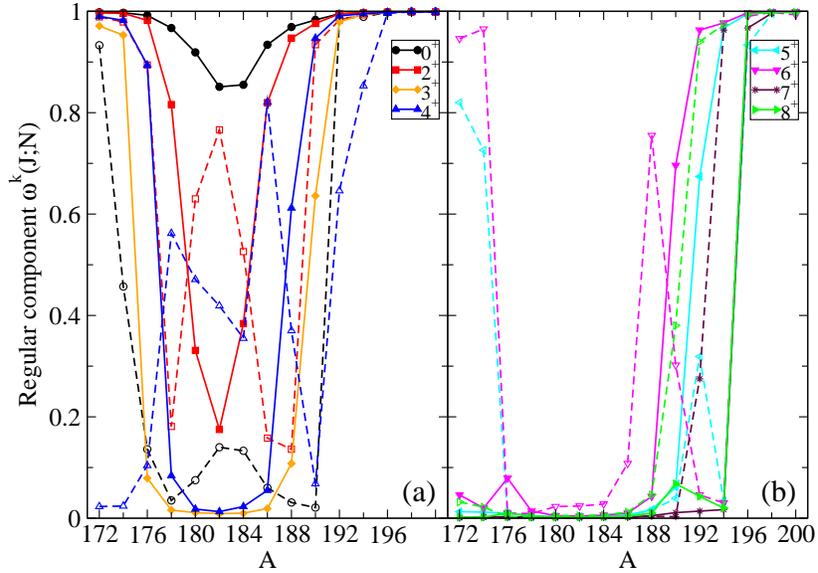}
  \caption{(Color online) Regular content of the two lowest-lying states
    for each $J$ value (full lines with
    closed symbols correspond with the first state while dashed lines with open
    symbols correspond with the second state) resulting from the IBM-CM
    calculation, as presented in figure \ref{fig-energ-comp}.}
  \label{fig-wf}
\end{figure}

\begin{figure}[hbt]
  \centering
  \includegraphics[width=.4\linewidth]{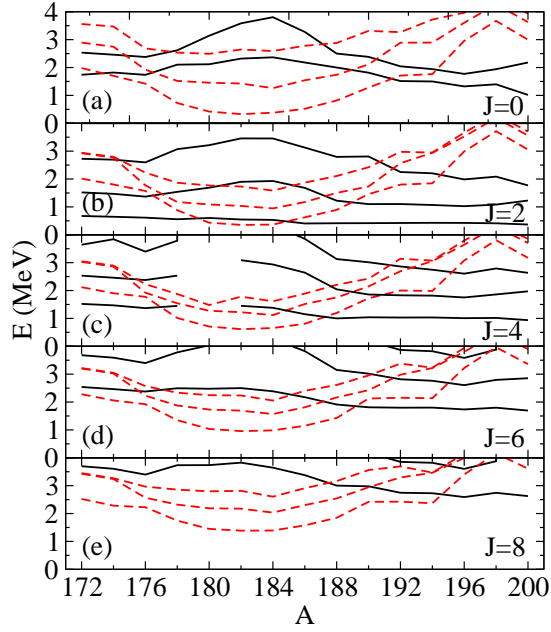}
  \caption{(Color online) Energy systematics for a set of selected states, separated by
    angular momentum. The full lines denote states with $w^k(J,N)>0.5$
    (mainly regular)
    and dashed lines for states with $w^k(J,N)<0.5$ (mainly intruder).}
  \label{fig-cross}
\end{figure}

In Fig.~\ref{fig-wf}, we present 
the weight of the wave functions contained within the $[N]$-boson subspace, defined as
the sum of the squared amplitudes $w^k(J,N) \equiv \sum_{i}\mid a^{k}_i(J;N)\mid ^2$, for
both the yrast states, $(k=1)$, and the $(k=2)$ states (the latter are
indicated with a dashed line) for spins $J^\pi=0^+, 2^+, 3^+, 4^+$ in panel (a) and
$J=5^+, 6^+, 7^+, 8^+$ in panel (b).  
The results exhibit an interesting behaviour, both as a function of
angular momentum $J$ and as a function of the changing mass number.
First, one notices the complementary behaviour of the $0_2^+$ state as compared to the $0_1^+$ state. This has important 
consequences for the study of the hindrance factor for $\alpha$ decay from the Pb ground state into the 
0$^+_{1,2}$ states in the Hg nuclei, as will be discussed in
Section~\ref{sec-other}. The $2_1^+$ and $2_2^+$ states also present 
the same complementary behaviour, interchanging their character near  
mid shell, however with the second $2^+$ state becoming of regular character by A=190 and onwards.
The $4^+_1$ state shows a very smooth behavior, almost fully symmetric around A=182, whereas the
second $4^+$ state shows some more complicated character, which is a consequence of the crossing
of a number of rather close-lying $4^+$ states when going from nucleus to nucleus (see Fig.~\ref{fig-cross}).
The $3_1^+$ state results to be mainly regular for the
lighter and heavier isotopes while mainly intruder at mid shell. 
The $3_2^+$ state is not depicted because a rather {\it
erratic} behaviour shows up which is due to the multiple crossing with other $3^+$
states. The higher-spin $5^+$, $6^+$, $7^+$, and $8^+$ states are
almost pure intruder states along the whole chain except for the
lightest and heaviest isotopes. Note that due to the construction of
the Hamiltonian for the heavier isotopes, we have imposed {\it by hand}
that the intruder states should be above $\sim 3$ MeV, forcing the wave
function of all these states to have a mainly regular structure. 
\begin{table}
\caption{Comparison of the experimental relative B(E2) values  with
    the IBM-CM Hamiltonian results. From left to right we give: isotope,
    transition, $\gamma$-ray energy, 
intensity of the transition, multipolarity,
experimental relative B(E2) value and the IBM-CM calculations. 
Data are taken from  
\cite{else11} for $^{180}$Hg , from \cite{rapi13} for $^{182,184}$Hg, \cite{gutt81} for $^{186,188}$Hg, and
\cite{korte91} for $^{190}$Hg. 
We use the expressions 
 $\mbox{B(E2)}= 100 \times \left (\frac{\mbox{I}_\gamma}{\mbox{I}_\gamma^{ref}}\right )\times
 \left(\frac{\mbox{E}_\gamma^{ref}}{\mbox{E}_\gamma }\right)^5$ and
 $\Delta(\mbox{B(E2)})= \mbox{B(E2)}\times\sqrt{\
 \left (\frac{\Delta(\mbox{I}_\gamma)}{\mbox{I}_\gamma}\right)^2+
 \left(\frac{\Delta(\mbox{I}_\gamma^{ref})}{\mbox{I}_\gamma^{ref}}\right)^2}$
 in order to extract the relative B(E2) values 
 and their corresponding relative errors.
} 
\begin{center}
\begin{ruledtabular}
\begin{tabular}{lllllll}
&Transition&E$_\gamma$(keV)&I$_\gamma$&Mult.&Exp.& IBM-CM\\
\hline 
$^{180}$Hg\footnotemark[1]
&$2_2^+\rightarrow 0_2^+$&181.8 &0.16(1)&E2&100 & 100 \\
        &$2_2^+\rightarrow 0_1^+$&601.6 &24.3(12)&E2&38(3) & 11 \\	
        &$4_1^+\rightarrow 2_2^+$&104.7 &1.4(4)&E2&100 & 100 \\	
        &$4_1^+\rightarrow 2_1^+$&272.0 &54.2(27)&E2&33(9) & 272 \\	
\hline 
$^{182}$Hg\footnotemark[2]
&$2_2^+\rightarrow 0_2^+$&213.0&1.4(5) &E2&730(260) & 554 \\
            &$2_2^+\rightarrow 0_1^+$&547.8&21.5(8) &E2&100 & 100 \\	
            &$4_2^+\rightarrow 2_2^+$&576.6&15.0(2) &E2&650(65) & 850 \\	
            &$4_2^+\rightarrow 2_1^+$&772.6&10(1) &E2&100 & 100 \\	
\hline 
$^{184}$Hg\footnotemark[2]&$2_2^+\rightarrow 0_2^+$&159.4&1.7(7) &E2&3500(1400) & 2350 \\
        &$2_2^+\rightarrow 0_1^+$&534.7& 20.7(7)&E2&100 & 100 \\	
        &$4_1^+\rightarrow 2_2^+$&118.8&0.5(3) &E2&250(150) & 45  \\	
        &$4_1^+\rightarrow 2_1^+$&287.0&16.7(3) &E2&100 & 100 \\	
        &$4_2^+\rightarrow 2_2^+$&552.0& 9.5(7)&E2&500(50) & 346 \\	
        &$4_2^+\rightarrow 2_1^+$&719.6& 7.1(4)&E2&100 & 100 \\	
\hline 
$^{186}$Hg\footnotemark[3]
&$2_2^+\rightarrow 0_2^+$&97 & &E2&$>10^5\pm 3\cdot 10^4$ & 24000 \\
        &$2_2^+\rightarrow 0_1^+$&621 & &E2&100 & 100 \\	
        &$4_1^+\rightarrow 2_2^+$&187 & &E2&210(90) & 303 \\	
        &$4_1^+\rightarrow 2_1^+$&402 & &E2&100 & 100 \\	
        &$4_2^+\rightarrow 2_2^+$&460 & &E2&110(20) & 130 \\	
        &$4_2^+\rightarrow 2_1^+$&675 & &E2&100 & 100 \\	
        &$6_2^+\rightarrow 4_2^+$&597 & &E2&2900(300) & 9500 \\	
        &$6_2^+\rightarrow 4_1^+$&870 & &E2&100 & 100 \\	
\hline 
$^{188}$Hg\footnotemark[3]&$4_2^+\rightarrow 2_2^+$&327 & &E2&8070(580) & 2170 \\	
        &$4_2^+\rightarrow 2_1^+$&795 & &E2&100 & 100 \\	
        &$6_1^+\rightarrow 4_2^+$&301 & &E2&270 (20) & 97 \\	
        &$6_1^+\rightarrow 4_1^+$&504 & &E2&100 & 100 \\	
        &$6_2^+\rightarrow 4_2^+$&569 & &E2&130(10) & 240 \\	
        &$6_2^+\rightarrow 4_1^+$&772 & &E2&100 & 100 \\	
\hline 
$^{190}$Hg\footnotemark[4]
&$2_4^+\rightarrow 0_2^+$&292 &  0.03(2)&E2&$6\cdot 10^4\pm 4\cdot 10^4$& 8470$^*$\\
        &$2_4^+\rightarrow 0_1^+$&1571 & 0.21(12) &E2&100 & 100 \\	
        &$4_2^+\rightarrow 2_4^+$&404 & 0.28(8)&E2&19000(6000) & 570$^*$\\	
        &$4_2^+\rightarrow 2_1^+$&1559 & 1.22(9)&E2&100 & 100 \\	
        &$6_2^+\rightarrow 4_2^+$&535 & 0.75(6)&E2&30700 (3500) & 7615$^*$ \\	
        &$6_2^+\rightarrow 4_1^+$&1468 & 0.38(3)&E2&100 & 100 \\	
\end{tabular}
\end{ruledtabular}
\end{center}
\label{tab-be2-1}

\footnotetext{}
\footnotetext[1]{Relative B(E2) values and error bars calculated from the data
($\gamma$ intensities with errors) given in \cite{else11}.}
\footnotetext[2]{Relative B(E2) values and error bars calculated from the data
($\gamma$ intensities with errors) given in \cite{rapi13}.}
\footnotetext[3]{Relative B(E2) values and error bars as given in
\cite{gutt81}.}
\footnotetext[4]{Relative B(E2) values and error bars as given in
  \cite{korte91}.}
\footnotetext{$^{*}$ The effective charges have been taken the same as
  the corresponding values obtained for $^{188}$Hg (see also table
  \ref{tab-fit-par-mix}).} 
\end{table}

Finally, to explain the sudden changes in the structure of the wave functions, 
we have to take into account that in most cases several states with identical angular
momentum remain approximately at the same excitation energy, but have
a different character. So, it can happen that close-lying states interchange character when passing 
from one isotope to the next one, resulting in an abrupt change in the wave
function content. This is depicted in Fig.~\ref{fig-cross} where we
show the energy spectra corresponding to the Hamiltonian given in Table
\ref{tab-fit-par-mix}, separating the different angular momenta,
as a matter of clarity, and distinguishing the character of the
states, using \textit{i.e.}, full lines for the regular states ($w^k>0.5$)
and dashed lines for
the intruder ones ($w^k<0.5$) (see panels (a) to (e) for
$J^\pi=0^+,...,8^+$, respectively). Note that the missing points correspond to regular states
with an excitation energy which is larger than $\sim 3$ MeV and do not appear within the scale
of the figure. We can use this figure to understand the observed behaviour as depicted in
Fig.~\ref{fig-wf}. As an illustration we consider the case of the $J^{\pi}=0^+$ states:
the first excited $0^+$ state corresponds to mainly a regular state
for A $\sim$ 172, then moves to become an intruder state 
up to A=190 and finally returns to become the regular ground state up
to A=200. The second excited $J^{\pi}=0^+$ state is an 
intruder state at the low-mass region with A $\sim$ 172, then moves to become the
first regular state up to A=176, proceeds to be characterized as the second intruder state up to A=188, as 
first regular state at A=190, as first intruder state (A=192-194) next, to end up as the second regular 
state (A=196-200).

\begin{figure}[hbt]
  \centering
  \includegraphics[width=.5\linewidth]{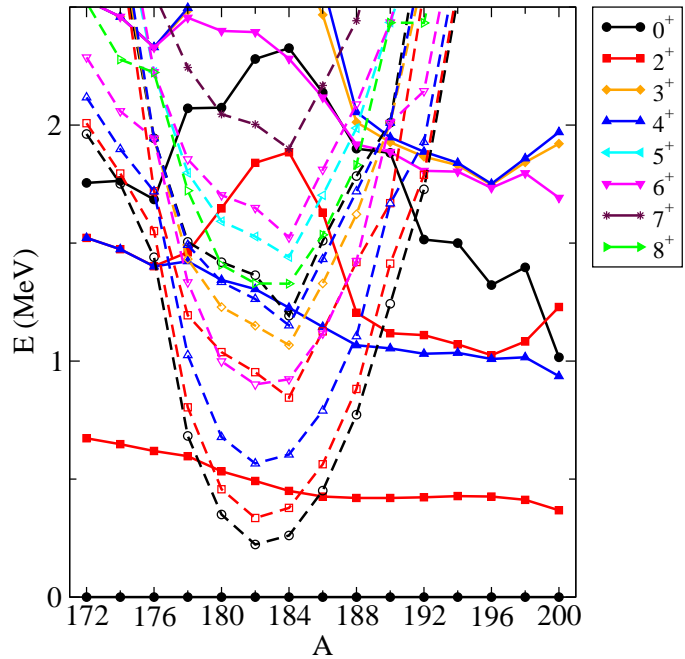}
  \caption{(Color online) Energy spectra for the IBM-CM Hamiltonian presented in table 
    \ref{tab-fit-par-mix}, switching off the mixing term. The two
    lowest-lying regular states and the
    lowest-lying intruder state for each of the angular momenta are shown (full lines with
    closed symbols for the regular states while dashed lines with open
    symbols are used for the intruder states).}
  \label{fig-ener-nomix}
\end{figure}

In order to understand more clearly the way the energy spectra have been affected by the mixing 
term, we recalculate the energy spectra using the Hamiltonian presented in 
table \ref{tab-fit-par-mix}, but now 
switching off the mixing term. The spectra are presented in
Fig.~\ref{fig-ener-nomix} 
where we show the lowest two regular and the lowest intruder
state for different angular momenta. One observes a rather flat behavior of
the energy for the regular states, but with an up sloping
tendence moving to the lighter isotopes. The energy of the intruder
states is smoothly decreasing up to neutron mid-shell (minimum occurs at N=102), where it starts
increasing again. 
This effect results mainly from the smooth change of the
Hamiltonian parameters when passing from isotope to isotope. 
A simultaneous analysis of figures \ref{fig-wf} and
\ref{fig-ener-nomix}, combined with the rules of a simple two-level mixing model, 
allows us to explain the sudden increase
of the regular content for all $J^{\pi}$ values at A=188. 
Inspecting Fig.~\ref{fig-ener-nomix}, one observes the close
approach of pairs of regular and intruder states with a given
angular momentum, especially in the region around A=188. 
The mixing term, coupling the regular ($N$) and intruder 
($N+2$) configurations, can now result in the
interchange of character between the states and therefore in the sudden
increase of the regular content  of the wave function. For states with $J>4$, the
effect is even more dramatic because the unperturbed energy of the intruder configuration
always lies below the unperturbed energy of the regular one and as a
consequence, the interchange in character
with the regular configuration at the point of closest approach is enhanced. Eventually, 
moving towards A=194, the unperturbed energy of the intruder configurations is moving up and
crosses the regular configurations. Therefore, as shown in Fig.~\ref{fig-wf}, from A=194  
onwards, the two lowest-lying states for each $J^{\pi}$ value have become regular (N-component,
mainly) states.

A most interesting decomposition of the wavefunction is obtained by first calculating the wavefunctions within the N subspace as
\begin{equation}
\Psi(l,JM)^{reg}_N = \sum_{i} c^{l}_i(J;N) \psi((sd)^{N}_{i};JM)~, 
\label{eq:wf:N}
\end{equation}
and likewise for the intruder (or N+2 subspace) as
\begin{equation}
\Psi(m,JM)^{int}_{N+2} = \sum_{j} c^{m}_j(J;N+2)\psi((sd)^{N+2}_{j};JM)~,
\label{eq:wf:N+2}
\end{equation}
defining an ``intermediate'' basis~\cite{helle05,helle08}. 
These wave functions correspond to the energy levels shown in Fig.\ref{fig-ener-nomix}.
This generates a set of bands within the 0p-0h and 2p-2h subspaces, 
corresponding to
the unperturbed bands that are extracted in schematic two-level phenomenological model calculations 
(as discussed in references ~\cite{drac88,duppen90,drac94,allat98,page03,drac04,lipska13}),
and indeed correspond to the unperturbed energy levels depicted in Fig.~\ref{fig-ener-nomix}.

\begin{figure}
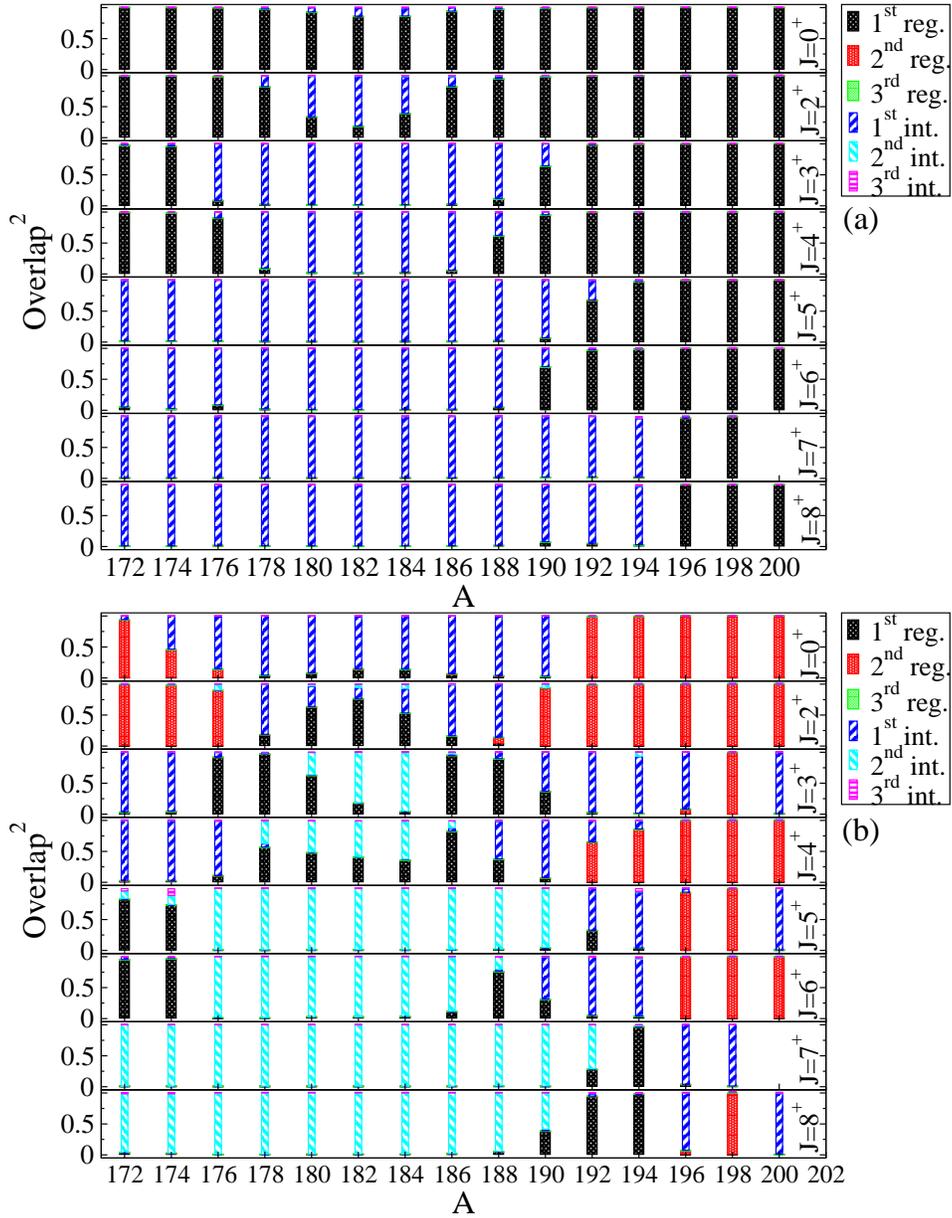

\includegraphics[width=0.7\textwidth]{overlap-int-reg-1.eps}\\
\includegraphics[width=0.7\textwidth]{overlap-int-reg-2.eps}%
\caption{(Color online) Overlap of the wave functions of Eq.~(\ref{eq:wf:U5}), with
  the wave functions describing the unperturbed basis  
Eq.~(\ref{eq:wf:N}) and Eq.~(\ref{eq:wf:N+2}). panel (a): overlaps
for first $0^+,2^+,3^+,4^+;5^+,6^+,7^+,8^+$ state,  
  panel (b): overlaps for the corresponding second state (see also text).}
\label{fig-overlap} 
\end{figure}

The overlaps $_{N}\langle l,JM \mid k,JM\rangle$ and $_{N+2}\langle m,JM \mid k,JM\rangle$
can then be expressed as,
\begin{equation}
_{N}\langle l,JM \mid k,JM\rangle=\sum_{i} a^{k}_i(J;N) c^{l}_i(J;N), 
\end{equation} 
and  
\begin{equation}
_{N+2}\langle m,JM \mid k,JM\rangle=\sum_{j} b^{k}_j(J;N+2) c^{m}_j(J;N+2),
\end{equation} 
(see expressions (\ref{eq:wf:N}) and (\ref{eq:wf:N+2})).
In Fig.~\ref{fig-overlap} we show these overlaps, but squared, where we restrict
ourselves to the first and second state  ($k=1,2$) with 
angular momentum J$^{\pi}$=0$^+$,2$^+$,3$^+$,4$^+$,5$^+$,6$^+$,7$^+$,8$^+$,
and give the overlaps with the lowest three bands within the
regular $(N)$ and intruder $(N+2)$ spaces ($l=1,2,3$ and $m=1,2,3$).  Since
these figures are given as a function of mass number, one obtains a graphical insight into the changing wave function
content. In particular, in the upper panel (a), corresponding to the first
state of each angular momentum, an
inverted parabola separating the regular and the intruder states as a
function of increasing angular momentum is clearly observed.  In the
lower panel (b),
the parabolic shape is also present but in this case it does not
separate so clearly into regular and intruder configurations. The
central region mainly corresponds to the second lowest intruder state while
the outer region corresponds to the second regular and the first
intruder state.

\section{Study of other observables: isotopic shifts and alpha-decay
  hindrance factors }
\label{sec-other}

\subsection{$\alpha$-decay hindrance factors}
\label{sec-alpha}

In the Pb-region, most interesting results were obtained when the
content of the nuclear wave functions was tested through
$\alpha$-decay measurements.  It was shown by Andreyev {\it et al.}~\cite{andrei00}
that $\alpha$-decay has been instrumental as a sensitive probe to prove the presence
of a triplet of $0^+$ states in $^{186}$Pb, each corresponding to a different
shape.

Wauters {\it et al.}~\cite{wauters94,wauters94a} carried out experiments
on the $\alpha$-decay from the Po, Pb and Hg nuclei to the Pb, Hg and Pt nuclei,
respectively, concentrating in particular on the N=104 mid-shell
region. $\alpha$ decay is a highly sensitive fingerprint, precisely
because an $\alpha$ particle is emitted in the decay,  a process which requires
the extraction of two protons and two neutrons from the initial nucleus.  
The comparison of s-wave {\it l}=0 $\alpha$-decay branches from a given
parent nucleus (the Pb $0^+$ ground state in the present situation) 
to $0^+$ states in the daughter nucleus (the Hg $0^+$ ground state and
excited $0^+$ states) is important in this respect.
\begin{figure}[hbt]
 \centering
 \epsfig{file=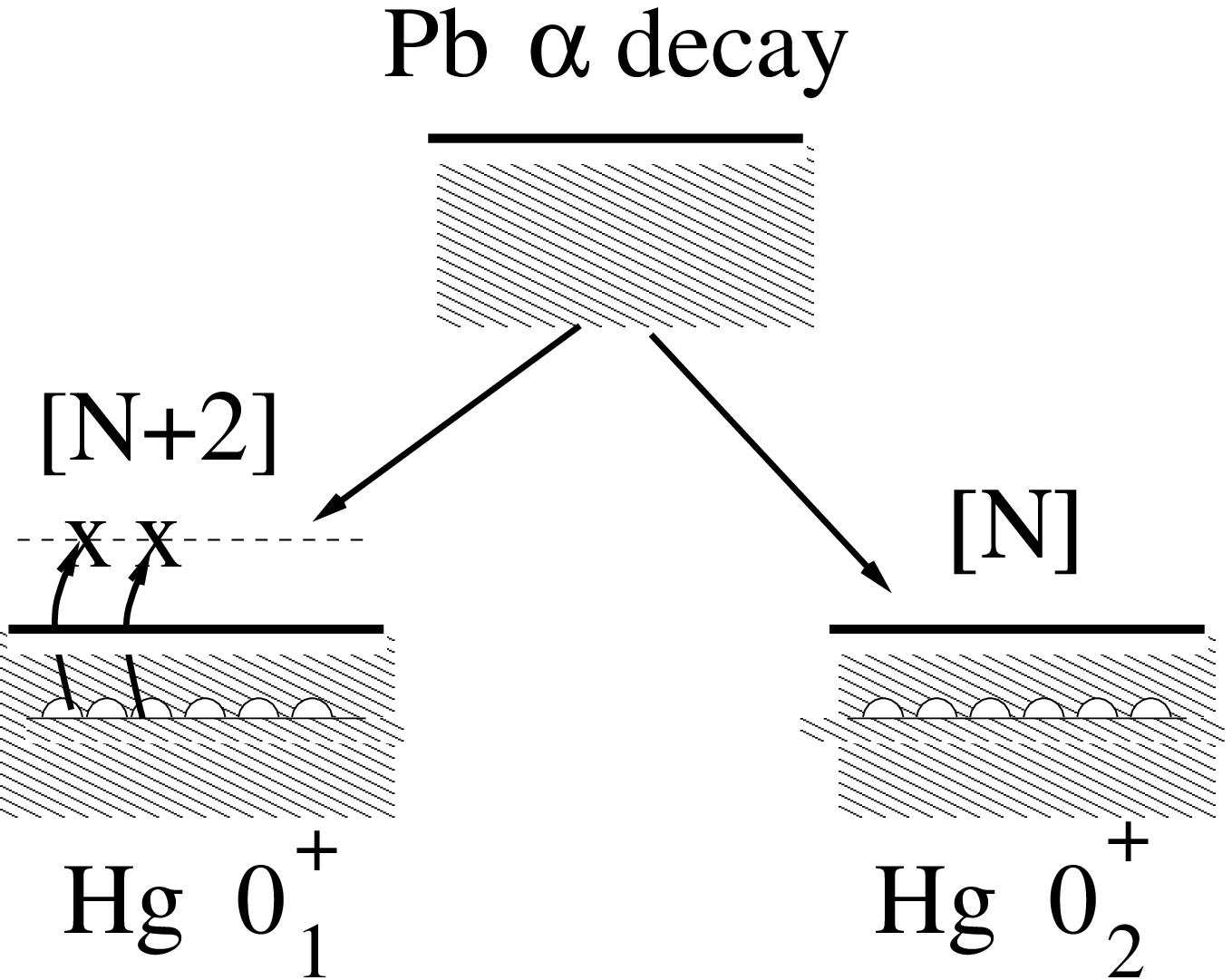,width=5cm} 
 \caption{A schematic view of the $\alpha$-decay proceeding from the 0$^+$ Pb ground
 state into the 0$^+_{1,2}$ states of the Hg nuclei.}
 \label{fig-pb-alpha}
\end{figure}
The reduced $\alpha$-decay widths themselves are very difficult to 
calculate on an absolute scale \cite{delion13}, but hindrance factors
clearly reflect possible changes amongst the wave functions
describing various $0^+$ states in a given daughter nucleus \cite{duppen00} well
(see \cite{Garc11} for the precise definition and applications to the Hg to Pt $\alpha$ decay hindrance
factor calculations as compared with the data).

These experiments indicated that, in the neutron mid-shell region, the $0^+$ ground-state in the Pb and Hg
nuclei is essentially consistent with a closed Z=82 core and a
two-proton hole configuration in the Z=82 core \cite{wauters94,wauters94a}
(see Fig.~\ref{fig-pb-alpha}). 
However, $\alpha$-decay feeding into the   
first-excited $0^+_2$ state exhibits a hindrance factor. 
The specific values of the hindrance factors are the adopted values as given in Nuclear
Data Sheets, starting from the original data \cite{wauters94,wauters94a}.
This is qualitatively in line with the results presented
in Fig.~\ref{fig-wf}, where the $0^+_1$ ground state mainly consists of the  
regular $[N]$ configuration, dropping to a value of $\approx 85 \%$ of the
$[N]$ component at A=$182,184$. The results of the present calculation indicate an almost
symmetric structure with respect to the mid-shell N=104 neutron number.
The important point here, as also stressed by Van Duppen and Huyse \cite{duppen00}, 
is the consistent picture that results when treating the Po, Pb, Hg, and Pt nuclei jointly. 
It turns out that the structure of the 
wavefunctions for the 0$^+_1$ and 0$^+_2$ states are consistent with the
wavefunctions extracted from both $\alpha$-decay hindrance factors and E0 transitions between
the ground and first excited 0$^+$ states \cite{delion95,richards97,wood99,joshi94}. 

\subsection{Isotopic shifts}
\label{sec-radii}

\begin{figure}[hbt]
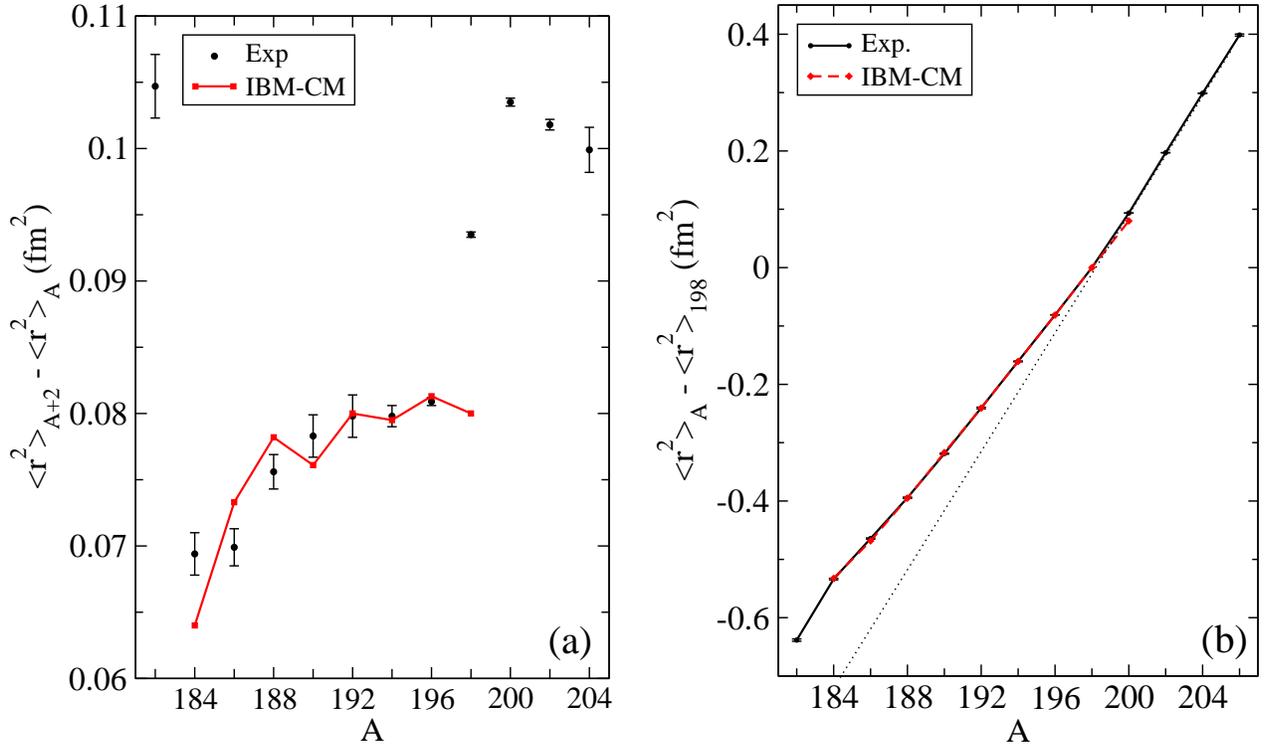

  \centering
  \includegraphics[width=.45\linewidth]{iso-shift.eps}%
~~~~
  \includegraphics[width=.45\linewidth]{exp-radius2.eps}
  \caption{(Color online) Panel (a): Isotopic shift for the Hg
    nuclei. Panel (b): Charge
    mean-square radii for the Hg nuclei. The data are taken from \cite{ulm86}.}
  \label{iso-shift}
\end{figure}

Experimental information about ground-state charge radii is also available
for both the even-even and odd-mass Hg nuclei.
Combined with similar data for the adjacent
Po, Pb, and Pt nuclei, as well as for the odd-mass Bi, Tl and Au nuclei, the systematic
variation of the charge radii supplies invaluable
information on the ground-state wave function \cite{otten89,kluge03}. 
We illustrate the relative changes defined as
$\Delta {\langle r^2 \rangle}_A \equiv \langle r^2 \rangle_{A+2}$ -$\langle
r^2 \rangle_{A}$ in Fig.~\ref{iso-shift} (left side) and
the overall behaviour of ${\langle r^2 \rangle}_A$ relative
to the radius at mass A=198 in Fig.~\ref{iso-shift} (right side). The
experimental data are 
taken from Ulm \textit{et al.}~\cite{ulm86}. 

To calculate the isotope shifts, we have used the standard IBM-CM expression for the nuclear radius
\begin{equation}
r^2=r_c^2+ \hat{P}^{\dag}_{N}(\gamma_N \hat{N}+ \beta_N
\hat{n}_d)\hat{P}_{N} + 
\hat{P}^{\dag}_{N+2}(\gamma_{N+2} \hat{N}+ \beta_{N+2} \hat{n}_d) \hat{P}_{N+2}.
\label{ibm-r2}
\end{equation} 
The four parameters appearing in this expression are adjusted
to the experimental data. Note that only the
experimental values past mid shell (A=184) are used. The resulting values
are $\gamma_N=-0.099$ fm$^2$, $\beta_N=0.004$ fm$^2$,
$\gamma_{N+2}=-0.059$ fm$^2$, and $\beta_{N+2}=0.013$ fm$^2$ and are
only valid for the second half of the shell.  

The panel (a) of Fig.~\ref{iso-shift} shows a relatively small variation of the
isotopic shift over the whole chain of the Hg isotopes (184 $\leq$ A
$\leq$ 204). However, two different regions are clearly marked: one
region from A=184 to A=196 in which a steady increase of the isotopic
shift is observed ending in a smooth stabilization at A = 196,
followed by the region from A=198 to A=204 where a sudden increase in
the isotopic shift occurs, suggesting the transition into a new
regime.  The mean-square charge radius exhibits (see panel (b) of Fig.~\ref{iso-shift}) 
a smooth decrease from A=204 down to A=184,
but exhibits systematic deviations from the linear trend, which is
marked with the dotted straight line. These data suggest that  no
major change in the ground-state structure appears along the whole
chain of the Hg isotopes, contrary to what is observed in the
odd-mass isotopes \cite{otten89,kluge03} (and cf.~Fig.~21 in
Ref.~\cite{heyde11}). 

The IBM-CM exhibits quantitative agreement up to A=196. For A=198,
a too small theoretical value is obtained as compared with the data,
where one observes a change to a significantly larger value of the
isotopic shift which tends to flatten at A= 204. It is a remarkable
fact that the overall variation of the isotopic shifts along the whole
isotopic Hg chain is very small $\approx 0.05$ fm$^2$, and this range
is fairly well reproduced by the IBM-CM calculations. This is in
contrast with the Pt isotopes in which the overall variation amounts
to $\approx 0.2$ fm$^2$ ~\cite{Garc11}. Therefore, the good
reproduction of this range confirms the calculated interplay between
the $[N]$ and $[N+2]$ contributions in the $0^+$ ground state wave
function along the whole chain of Hg isotopes.

\section{The connection of the IBM-CM and the mean field studies}
\label{sec-meanfield}

\subsection{Mean-field studies: phenomenological potentials and self-consistent methods}

The Hg nuclei have been studied using mean-field methods emphasizing the intrinsic structure of
atomic nuclei \cite{bender03}. A more phenomenological approach, within the same spirit,
has been used to study the Hg nuclei (and nuclei in the Pb region taken more generally) starting from a deformed
Woods-Saxon (DWS) potential as an approximation to a deformed mean-field. We stress the fact
(see discussion further on) that the intrinsic property such as an oblate and prolate shape, or,
more generally, a shape defined over the $\beta,\gamma$ domain, is not an observable and its
use to confront them with data only serves as a qualitative guide. A clean separation of shapes,
which has been shown through extensive experimental studies in the Pb nuclei, leads to good
evidence for the coexistence of spherical, oblate and prolate shapes. However, moving
away from the Pb nuclei, into the Hg, Pt and Po, Rn nuclei, much stronger mixing is expected
and, as such, a discussion starting from the intrinsic frame, at lowest order, can only be a
starting point.

Concentrating on the Hg nuclei, the total energy has been calculated starting 
from a deformed Woods-Saxon potential (DWS) \cite{bengt87,naza93}. The results
give rise to spherical shapes up to A=176, followed by a region where a slightly oblate
shape ($\beta \sim$ -0.13 to -0.10) and a prolate shape ($\beta \sim$ 0.20 to 0.25) for 178 $\leq$ A $\leq$ 188, 
coexist, changing into oblate shapes only from mass A=190 onwards, ending  in a spherical shape at A=200. 
These studies are restricted to axial systems. In a number of papers, the obvious point is made
to associate the calculated prolate-oblate energy difference $\Delta E_{po} = E_{prolate}
- E_{oblate}$ with the experimental energy difference $\Delta E_x= E_x(0^+_2) - E_x(0^+_1$).
In those cases when strong mixing is involved, however, it can become unsafe to 
use this approach in order to decide on a given character of ``observed'' states as corresponding
to prolate and/or oblate states.

The early mean-field calculations of Girod and Reinhard \cite{girod82}, using axial
quadrupole deformation presented very much the same outcome with respect to shape coexistence
in the interval 180 $\leq$ A $\leq$ 188. Covering the full $\beta,\gamma$ plane,
Delaroche \textit{et al.} \cite{dela94} showed the appearance of shape coexistence in the
184 $\leq$ A $\leq$ 188 isotones, with clear indications for triaxal bands in A=188.

Relativistic-mean-field calculations, with specific application to the Hg nuclei, using the
NL1 effective interaction \cite{patra94,yoshida94} resulted in both serious overbinding for these nuclei and moreover
indicated that the lowest energy in the 178$\leq$A$\leq$188  mass region was associated with a prolate
shape, contrary to the non-relativistic calculations \cite{girod82,dela94}. Using a different
treatment of the pairing energy, this time the lowest energy obtained corresponded with an oblate shape \cite{yoshida97}.
The problem here is that on the scale of total binding energies, differences with the data as large
as $\sim$10 MeV resulted, with a value for $\Delta E_{po}$ of $\sim$0.5 MeV only, which makes this result very sensitive
to the precise prescription used. Niksic \textit {et al.} \cite{niksic02} made a thorough study aiming
to construct an effective interaction, called NL-SC. The constraint was to reproduce as well as possible
the experimental gap in the proton single-particle spectrum, as this quantity is of major importance
in deriving correctly the energy cost to create np-mh excitations across the Z=82 closed shell. With this
force, called NL-SC, both the total binding energy and charge radii for the Hg nuclei were reproduced
rather well. Moreover, as was the case with the DWS calculations, the oblate minimum becomes the lowest in
the 178$\leq$A$\leq$188 region.

It is interesting to point out that most recent mean-field total energy calculations,
either starting from the Gogny D1M force \cite{nomura12a} or from the Skyrme SLy6 force
\cite{yao13} result in the prolate energy minimum being the lowest (even after projecting on
angular momentum J in the latter case) from $^{180}$Hg up to $^{186}$Hg (with the oblate and prolate
minima becoming very close at mass A=186 in the former). In the latter calculation,
Yao \textit{et al.}~\cite{yao13} have moved beyond the mean-field, using the constraint of axial symmetry,
and calculated the observables so as to confront the theoretical approach with the genuine data set
(which they did for the Pb and Po nuclei, as well). In both calculations, the oblate minimum remains
the only one, once having reached A=190 and beyond. What becomes clear
is that collective dynamical correlations 
determine the final outcome of the nuclear properties, definitely in the case of the Hg
nuclei where the oblate and prolate minima in the region  178$\leq$A$\leq$188 are quite close, resulting
in a shallow energy surface \cite{yao13} along the triaxial quadrupole deformation, indicating the need
for GCM calculations covering the full $\beta,\gamma$ plane.    
\begin{figure}
\includegraphics[width=0.3\textwidth]{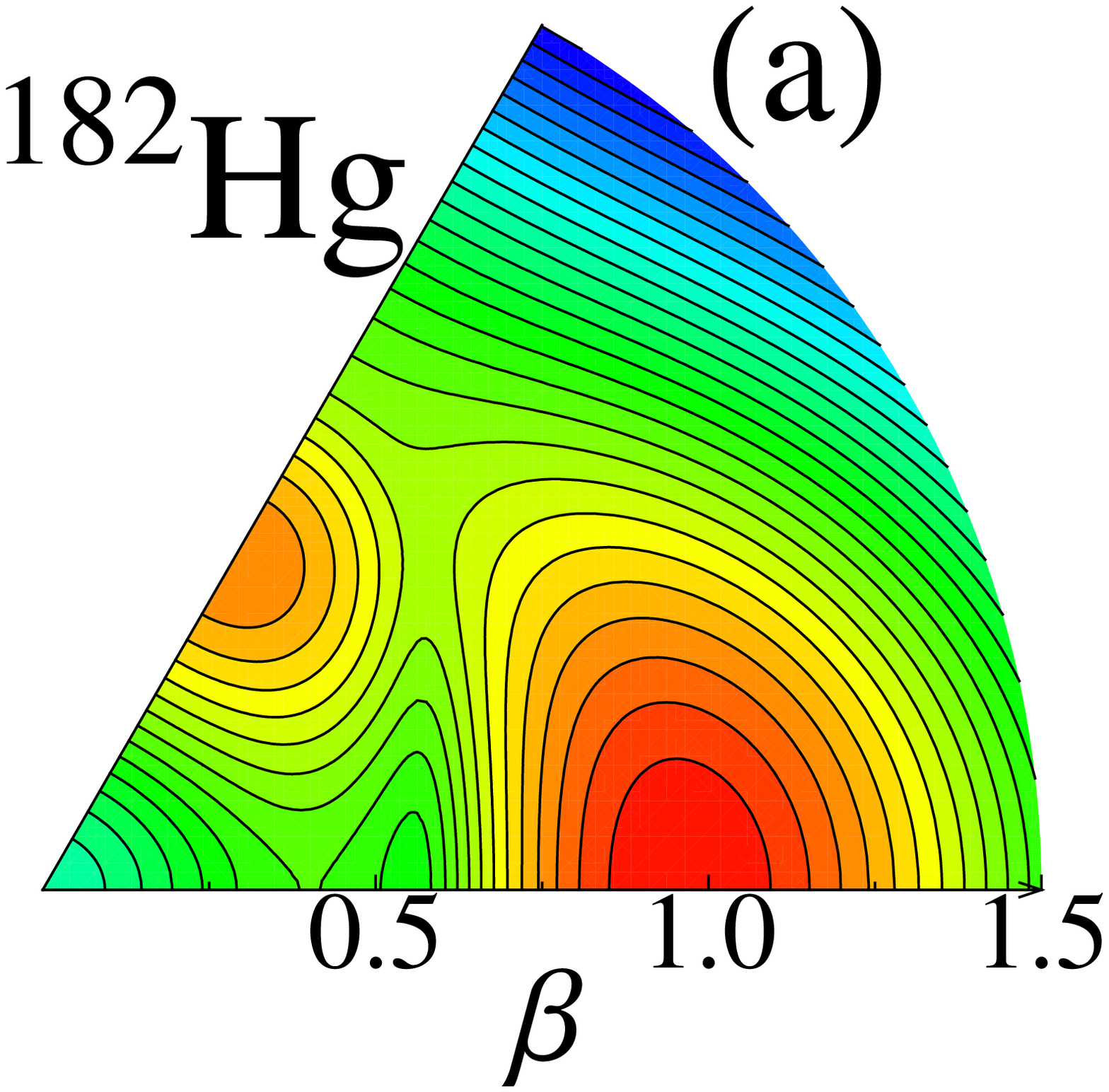}%
\includegraphics[width=0.3\textwidth]{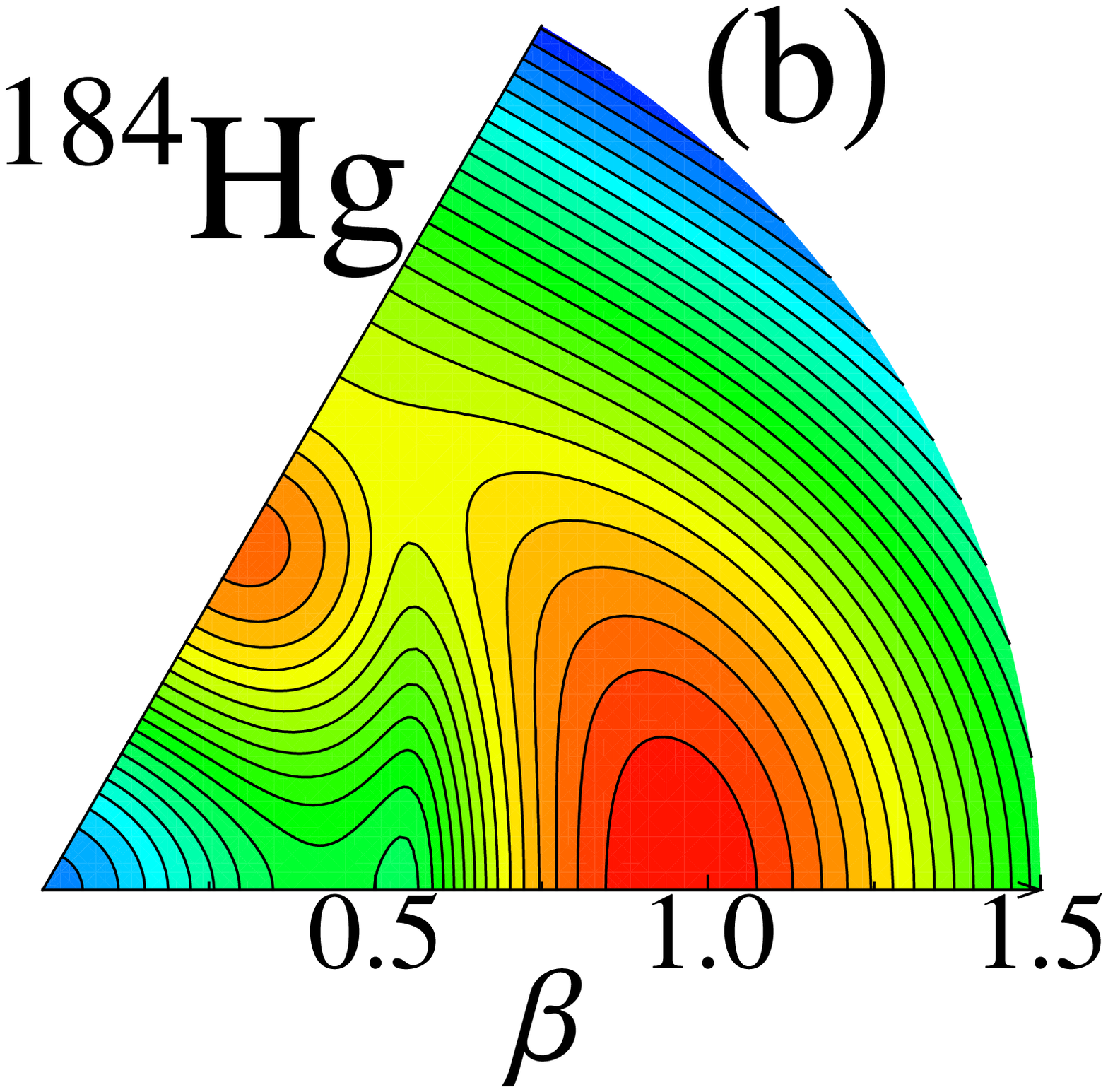}%
\includegraphics[width=0.3\textwidth]{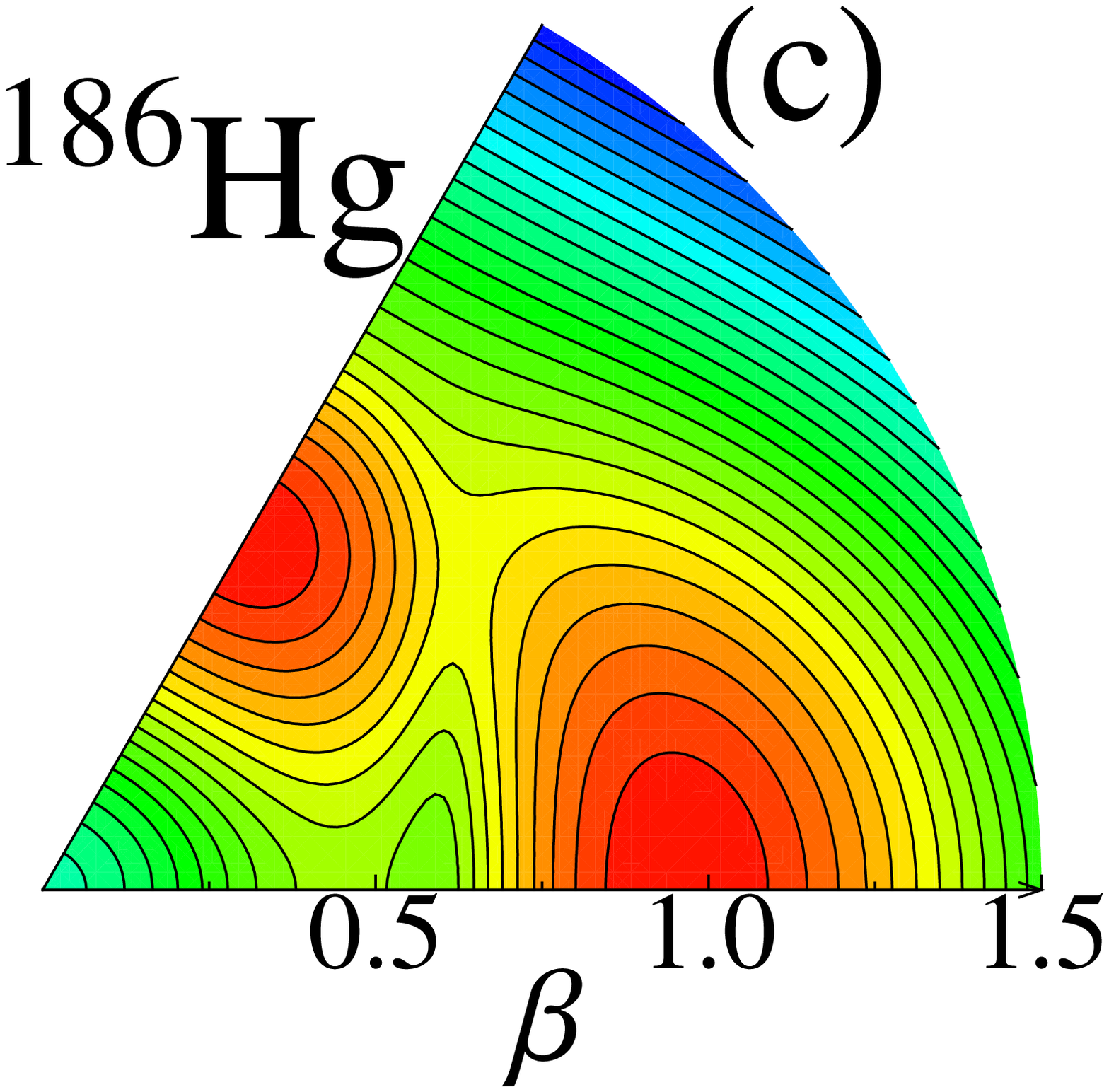}
\caption{(Color online) Matrix coherent-state calculation for $^{182-186}$Hg,
  corresponding with the present IBM-CM Hamiltonian (table
  \ref{tab-fit-par-mix}). The energy spacing  between adjacent contour lines equals $100$ keV and the
  deepest energy minimum is set to zero, corresponding to the red color.}
\label{fig_ibm_ener_surph}
\end{figure} 

\subsection{Mean-field approximation to the IBM: the energy surfaces}
\label{sec-ibm-mf}

A geometric interpretation of the IBM can be constructed using the
intrinsic state formalism, as proposed in the 1980's by Bohr and Mottelson \cite{Bohr80},
Ginocchio {\it  et al.}~\cite{gino80} and by Dieperink {\it et
  al.}~\cite{diep80a,diep80b}, based on the concept of coset spaces
\cite{gil74}.  
This provides a simple way to connect with the intrinsic geometric 
mean-field properties of the model and therefore to obtain a simple picture
of the shape of the nuclei. 
 
To define the intrinsic state, one
assumes that the dynamical behavior of the system can be described using
independent bosons (``dressed bosons'') moving in an average
field \cite{Duke84}. The ground state of the system is a condensate
$|N; \beta,\gamma \rangle$ of bosons, occupying the lowest--energy phonon state,
$\Gamma^\dag_c$,  
\begin{equation}
\label{GS}
|N; \beta,\gamma   \rangle = {1 \over \sqrt{N!}} (\Gamma^\dagger_c)^N | 0 \rangle,
\end{equation}
where
\begin{equation}
\label{bc}
\Gamma^\dagger_c = {1 \over \sqrt{1+\beta^2}} \left (s^\dagger + \beta
\cos     \gamma          \,d^\dagger_0          +{1\over\sqrt{2}}\beta
\sin\gamma\,(d^\dagger_2+d^\dagger_{-2}) \right) ,
\end{equation}
and
$\beta$ and $\gamma$ are variational parameters related with the shape
variables in the geometrical collective model~\cite{bomot75}. The expectation value
of the Hamiltonian in the intrinsic  state (\ref{GS}) provides  
the energy surface of the system, $E(N,\beta,\gamma)=\langle N; \beta,\gamma 
|\hat H| N; \beta,\gamma  \rangle$ and the values of $\beta$ 
and $\gamma$, which minimize the
expectation value of the energy, represent the shape of the
nucleus. The energy surface obtained in this way is equivalent, up to
a scale factor, to the one derived from mean field theory ~\cite{bender03}.
The IBM value of $\gamma$ is directly comparable with the mean field value
\cite{nomura08,nomura10,nomura11a,nomura11b,nomura13,nomura12a}, 
while $\beta$ should be rescaled \cite{gino80}.  

To analyze the nuclear geometry in the case of IBM-CM, the 
intrinsic state formalism should be extended. This extension
was recently proposed by Frank {\it et al.,} introducing a matrix coherent-state method
\cite{Frank02,Frank04,Frank06,Mora08} that allows to describe shape coexistence in a geometric way.   

The way to proceed is to define a model space with the 
states $|N; \beta,\gamma  \rangle$, $|\ N +2; \beta,\gamma  \rangle$ 
and to diagonalize the Hamiltonian (\ref{eq:ibmhamiltonian}). So, one needs to
construct the $2\times 2$ matrix:
\begin{equation}
H_{CM}=\left (
\begin{array}{cc}
E_N(N,\beta,\gamma)& \Omega(\beta)\\
\Omega(\beta)& E_{N+2}(N+2,\beta,\gamma)+\Delta^{N+2}
\end{array}
\right ) ,
\label{surf-cm}
\end{equation}
in which 
$E(N,\beta,\gamma)=\langle N; \beta,\gamma  |\hat H |N; \beta,\gamma  \rangle$,   
$E(N+2,\beta,\gamma)=\langle  N+2; \beta, \gamma |\hat H | N+2; \beta, \gamma \rangle$, and
$\Omega(\beta)= \langle N; \beta,\gamma  |\hat H | N+2; \beta,\gamma
\rangle$. 
The terms $E_N(N,\beta,\gamma)$ and $E_{N+2}(N+2,\beta,\gamma)$ only contain
the $N$ and the $N+2$ contributions of the Hamiltonian (\ref{eq:ibmhamiltonian}),
respectively, while $\Omega(\beta)$ corresponds to the 
matrix element of the mixing
term $\hat{V}_{\rm mix}^{N,N+2}$. Note that $\Omega(\beta)$ only
depends on $\beta$, while  $E_N(N,\beta,\gamma)$  and
$E_{N+2}(N+2,\beta,\gamma)$ depend on $\beta$ and $\gamma$. The
explicit expressions of these matrix elements (see ref.~\cite{Mora08}) are: 
\begin{eqnarray}
\nonumber
E_i(N_i,\beta,\gamma)&=&(\varepsilon_i+6\kappa'_i)\frac{N_i\beta^2}{1+\beta^2}+\kappa_i
\left(
\frac{N_i}{1+\beta^2}(5+(1+\chi_i^2)\beta^2)+\frac{N_i(N_i-1)}{(1+\beta^2)^2}\right
.\\
&\times&\left.\big(\frac{2}{7}\chi_i^2\beta^4-4\sqrt{\frac{2}{7}}
\chi_i\beta^3\cos{3\gamma}+4\beta^2\big)\right),\\
\Omega(\beta)&=&\frac{\sqrt{(N_i+2)(N_i+1)}}{1+\beta^2}\left
    (w_0^{N,N+2} +w_2^{N,N+2}\frac{\beta^2}{\sqrt{5}}\right). 
\end{eqnarray} 
To obtain the energy surface one has to
diagonalize (\ref{surf-cm}) and to consider the lowest eigenvalue. The
meaning of the higher eigenvalue is not yet fully understood and will not
be used along this paper. 

Since its introduction, the intrinsic state formalism for IBM-CM
Hamiltonians has been used in very few cases
\cite{Frank06,Mora08,nomura12a,gar13_unpu}, 
however it can provide complementary information to 
the results from the IBM-CM in
the laboratory frame, in particular about the energy
surface and the shape of the nucleus. 
This is very useful in order to compare the energy surface, here starting from a lab frame formulation (the IBM),
with the total energy calculated using self-consistent HFB mean field methods.
\begin{figure}
\includegraphics[width=0.6\textwidth]{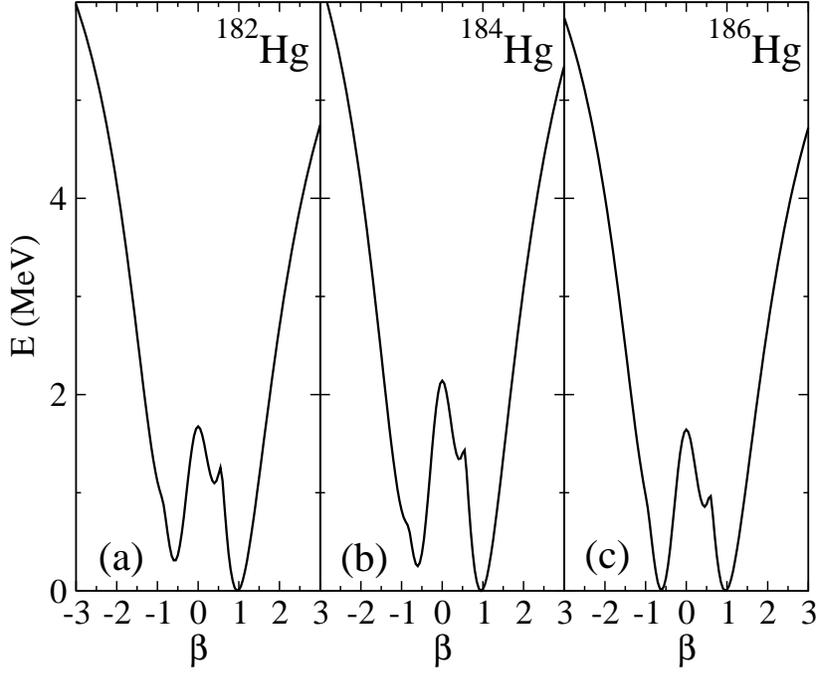}
\caption{Axial-symmetric energy curves for $^{182-186}$Hg using the matrix
  coherent-state calculation.}
\label{fig_ibm_ener_curves}
\end{figure} 

\begin{figure}
\includegraphics[width=0.4\textwidth]{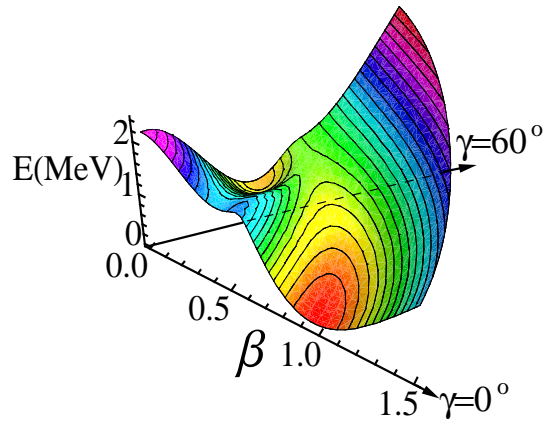}
\caption{(Color online) Three dimension energy plot for $^{184}$Hg using the matrix
  coherent-state calculation.}
\label{fig_3D_184}
\end{figure}

We have calculated the energy surfaces of the whole
chain of Hg isotopes, $^{172-200}$Hg, however a full analysis will be
presented elsewhere~\cite{gar13_unpu}. In this section, we focus on the
particular cases of $^{182}$Hg, $^{184}$Hg, and $^{186}$Hg which are at the mid
shell and have been analyzed in detail in \cite{yao13} through a beyond
mean-field HFB calculation. In Fig.~\ref{fig_ibm_ener_surph} (panels
(a), (b), and (c)) we present
the energy surface of $^{182-186}$Hg in the $\beta-\gamma$ plane, while in
Fig.~\ref{fig_ibm_ener_curves} we depict the axial-symmetric energy curves for
the three isotopes. Moreover, as a complementary view, we depict
  in Fig.~\ref{fig_3D_184} the three-dimensional energy plot of
  $^{184}$Hg. These figures show that the region around the minima
is rather shallow in both the $\beta$ and $\gamma$ directions. Inspecting the energy curves along the axial line only,
the three nuclei correspond to prolate shapes although  in
$^{186}$Hg the minima are almost degenerate. However, in these three cases there
are prolate and oblate minima connected by a path characterized by a
much lower barrier 
(see Fig.~\ref{fig_3D_184}). In Ref.~\cite{yao13}, using the Skyrme
Sly6 force, the HFB energy surface in 
the $\beta-\gamma$ plane for $^{184}$Hg is shown. That figure is equivalent, up
to a scale factor in $\beta$, to the IBM-CM results. Moreover the
axial energy curves are also very similar to the IBM-CM results. In
both approaches the three nuclei are prolate, up to the mean-field
level, but due to the flat behavior of the energy surface there is not
a clear-cut
preference for a dominant oblate nor prolate shape characterizing these nuclei.
Indeed, as can be observed in Fig.~\ref{fig_ibm_ener_curves} (panels
(a), (b), and (c)), the
prolate and the oblate minima - which are real minima and not saddle
points - are almost degenerate in energy and although there is a large barrier
at the spherical shape, a rather flat path connects
both minima going through the triaxial region. Therefore, in order to make the step towards the
observable properties, there is the need to include the quadrupole
collective dynamics, either restricting to axial symmetry or extending into the full triaxial
$\beta-\gamma$ plane, and go beyond mean field.

\subsection{The quadrupole invariants}
\label{sec-q-invariants}

The IBM can provide us with both the energy spectrum, the corresponding wave functions as well
as all derived observables (B(E2)'s, quadrupole moments, radii,...), working within the lab frame, as well as the
corresponding mean-field energy surface, defining a nuclear shape over the $\beta,\gamma$ intrinsic
frame.
Even though the shape of the nucleus is not an experimental observable, it is still possible to extract
from the data direct information about various moments characterizing the nuclear shape corresponding
with a given eigenstate. Using Coulomb excitation, it is possible to extract the most important diagonal
and non-diagonal quadrupole and octupole matrix elements, including their relative signs and, in a model
independent way, extract information about nuclear deformation  
as shown by Kumar, Cline and co-workers \cite{kumar72,Cline86,Wu96,clement07,sreb11}. 

The essential point is the introduction of an ``equivalent ellipsoid'' view of a given nucleus ~\cite{kumar72}
corresponding to a uniformly charged ellipsoid with the same charge, $\langle r^2 \rangle$, and 
E2 moments as the nucleus characterized by a specific eigenstate \cite{kumar72,wood12}.

From the theoretical point of view the nuclear shape can be calculated using the quadrupole shape
invariants. They correspond to:
\begin{eqnarray} 
q_{2,i}&=&\sqrt{5} \langle 0_i^+| [\hat{Q} \times \hat{Q} ]^{(0)}|0_i^+
\rangle,\\
q_{3,i}&=&-\sqrt{\frac{35}{2}} \langle 0_i^+[ \hat{Q} \times \hat{Q} \times \hat{Q}]^{(0)}|0_i^+
\rangle.
\label{q_invariant1}
\end{eqnarray} 
Note that these observables can be calculated for any $0_i^+$ state but
to simplify the notation we do not write explicitly the index $i$. 
For the triaxial rigid rotor model~\cite{wood09} they are directly connected with the
deformation parameters,
\begin{eqnarray} 
q_2 &=& q^2\label{q_invariant2b},\\
q_3 &=& q^3 \cos{3~\delta}, 
\label{q_invariant2}
\end{eqnarray} 
where $q$ denotes the nuclear intrinsic quadrupole moment and $\delta$ the triaxial
degree of freedom,
\begin{eqnarray} 
q&=&\sqrt{ q^2 },\\
\delta &=&\frac{60}{\pi} \arccos{\frac{q_3}{q_2^{3/2}}}.
\label{q_invariant3}
\end{eqnarray} 
The value of $\delta$ is equivalent, up to first-order
  approximation, to the value of $\gamma$ appearing in section \ref{sec-ibm-mf}
  (see Ref.~\cite{Srebrny06} for further details).

\begin{table}
\begin{ruledtabular}
\caption{Quadrupole shape invariants for the $^{180-200}$ Hg
  isotopes. Experimental values are taken from \cite{lipska13}.} 
\label{tab-q-invariant}
\begin{tabular}{cccccccc}
Isotope & State &\multicolumn{2}{c}{$ q^2 $ (e$^2$b$^2$)} & 
\multicolumn{2}{c}{ $\cos 3~\delta$ } & 
\multicolumn{2}{c}{$\delta$
  (deg)} \\
      &  & Theo. & Exp. & Theo. & Exp. & Theo. & Exp. \\
\hline
$180$ & $0_1^+$ &$1.83$  & - &$-0.01$  & -  & $30.3$ & - \\
      & $0_2^+$ &$6.54$  & - &$0.63$   & -  & $17.0$ & - \\
$182$ & $0_1^+$ & $2.00$ & $2.02(_{-15}^{+16}) $  & $0.23$ & $-1.33(_{-87}^{+109}) $ & $25.5$ & $(35,60)$ \\
      & $0_2^+$ & $6.19$ &$9.7(_{-51}^{+74}) $  & $0.65$ & $0.58(_{-61}^{+98}) $ & $16.6$ & $18(_{-18}^{+13})$ \\
$184$ & $0_1^+$ & $2.08$ &$1.66(12) $   & $-0.08$& $-1.04(_{-46}^{+42})$ &$31.5$  & $(43,60)$ \\
      & $0_2^+$ & $5.40$ &$13.9(_{-69}^{+97})$    & $0.54$ & $-0.34(_{-36}^{+34})$ & $19.2$ & $37(_{-7}^{+8})$ \\
$186$ & $0_1^+$ & $1.45$ & $1.56(_{-25}^{+23}) $  & $-0.13$& - &$32.6$  & - \\
      & $0_2^+$ & $6.07$ & - &$ 0.60$ & - &$17.7$  & - \\
$188$ & $0_1^+$ & $1.77$ &   &$-0.20$ &  &$33.9$  &  \\
      & $0_2^+$ &$5.62$  & - &$0.59$   & -  & $18.0$ & - \\
$190$ & $0_1^+$ &$1.48^*$  & - &$-0.19^*$  & -  & $33.7^*$ & - \\
      & $0_2^+$ &$4.78^*$  & - &$-0.03^*$  & -  & $30.7^*$ & - \\
$192$ & $0_1^+$ &$1.11^*$  & - &$-0.24^*$  & -  & $34.6^*$ & - \\
      & $0_2^+$ &$0.87^*$  & - &$-0.38^*$  & -  & $37.5^*$ & - \\
$194$ & $0_1^+$ &$0.91^*$  & - &$-0.16^*$  & -  & $33.1^*$ & - \\
      & $0_2^+$ &$0.67^*$  & - &$-0.19^*$  & -  & $33.7^*$ & - \\
$196$ & $0_1^+$ &$1.14^*$  & - &$-0.13^*$  & -  & $32.5^*$ & - \\
      & $0_2^+$ &$0.74$  & - &$-0.39$  & -  & $37.7$ & - \\
$198$ & $0_1^+$ &$0.97$  & - &$-0.29$  & -  & $35.5$ & - \\
      & $0_2^+$ &$0.57$  & - &$-0.57$  & -  & $41.7$ & - \\
$200$ & $0_1^+$ &$0.87$  & - &$-0.79$  & -  & $47.3$ & - \\
      & $0_2^+$ &$0.64$  & - &$-0.87$  & -  & $50.3$ & - \\
\end{tabular}
\end{ruledtabular}
\footnotetext{$^*$ The effective charges have been taken the same as
  the corresponding 
  values obtained for $^{188}$Hg, except for $^{196}$Hg where the ratio
  of e$_{N+2}$/e$_N$ was imposed to have
  the same value as in $^{188}$Hg (see also table \ref{tab-fit-par-mix}).}
\end{table}

To calculate analytically the quadrupole shape invariants characterizing the nucleus in its
ground-state and low-lying excited states, it is necessary to resort to a closure relation,
$\textbf{1}=\sum_{J,i,M}| J_i M\rangle \langle J_i M|$,
\begin{eqnarray} 
\label{q_invariant4}
q_{2,i} &=&\sum_r \langle 0_i^+|| \hat{Q}|| 2_r^+\rangle  \langle 2_r^+||\hat{Q} ||0_i^+ \rangle,  \\
q_{3,i} &=&-\sqrt{\frac{7}{10}}\sum_{r,s} 
\langle 0_i^+|| \hat{Q}||2_r^+\rangle  
\langle 2_r^+||\hat{Q} ||2_s^+ \rangle 
\langle 2_s^+||\hat{Q} ||0_i^+ \rangle.  
\label{q_invariant4b}
\end{eqnarray} 
Calculations of quadrupole shape invariants were carried out before
within the framework of the IBM by Jolos {\it et al.}~\cite{Jolo97}
and later by Werner and coworkers in Ref.~\cite{Werner00}.

A comparison with the experimental values can be carried out whenever a large enough set of reduced E2 matrix elements can 
be extracted from, \textit{e.g.}, Coulomb excitation experiments (see \cite{Cline86,Wu96} for a discussion on
the determination of the relative signs of these reduced E2 matrix elements). Such a comparison constitutes a very
stringent test for the theoretical model and, at the same time, provides
a clear picture of the nuclear shape. In table \ref{tab-q-invariant} 
we provide the theoretical values for the quadrupole shape invariants for all the
nuclei where it has been possible to fix the effective charges and we
also compare with the available experimental data. 

In table \ref{tab-q-invariant}, we compare our results with recent
Coulomb excitation experiments of $^{182-188}$Hg at REX-ISOLDE and
Miniball, allowing to extract a useful set of reduced E2 matrix
elements ~\cite{lipska13}.  It turns out that our IBM-CM calculations
indeed give rise to values of 
$q^2$ 
that differ by a
factor of $\approx$ 3 between the $0^+_1$ and $0^+_2$ states. The
calculated values are particularly stable, independent of the mass
number A=182-188 whereas the experimental data seem to indicate an
increasing deformation value with increasing A.  Evaluating the
invariant 
$q^3 \cos(3\delta) $ 
requires a lot more
matrix elements imposing restrictions in the summation over the
possible intermediate states. This will be particularly important for
the $0^+_2$ state as one can expect still important matrix elements
connecting to higher-lying $2^+_i$ states (even up to $i=4$). Here, our
calculated values do not show a particular correlation with the
experimentally extracted values for $\langle \cos(3\delta) \rangle$.

We have investigated the convergence in the summation over
the intermediate basis and found that in the four nuclei studied, \textit{i.e.},
$^{182-188}$Hg, the
summation is particularly sensitive to the number of $2^+_i$ states
included. To exemplify this effect, we focus on the nucleus
$^{184}$Hg. According to Eqs.~(\ref{q_invariant4})-(\ref{q_invariant4b}), used 
to calculate the value of $q_{2,i}$ and  $q_{3,i}$, it is not
known how many matrix elements will have to be included to reach a
convergent result. One could expect 
the matrix elements to rapidly fall (becoming vanishingly small)
with increasing excitation energy of the $2^+_i$ states involved. 
Therefore, one could expect that the sum can safely be truncated,
retaining only very few terms. 
We show the effect of the number of states on the deformation
parameters in table~\ref{tab-serie-q2-cos} concerning the variation of 
$q^2$ 
and of 
$\cos{3\delta}$ 
for the $0_1^+$ and $0_2^+$ states, as a function of the number of
$2^+_i$ states in the sum. One notices 
that in order to obtain a stable value for 
$\cos{3\delta}$, 
one needs to include more $2^+$ states than is the case for 
$q^2$. 
One also notices that 
convergence sets in faster for the $0^+_1$ state as compared to the $0^+_2$,
at least in the case of  
$q^2$. 
Particularly striking is the change in sign of 
$\cos{3\delta}$  
for the $0_1^+$ state, when increasing the number of intermediate $2^+$
states from one to two (in the calculation) and, considering up to
five intermediate $2^+$ states, and oscillation sets in.

\begin{table}
\begin{ruledtabular}
\caption{Calculated value of 
  $q^2$ and $\cos{3\delta}$ 
  for the $0_1^+$ and $0_2^+$ states in $^{184}$Hg, as
  a function of the number of $2^+$ states included, up to five. The
  exact value is also given.} 
\label{tab-serie-q2-cos}
\begin{tabular}{ccccccc}
\multicolumn{7}{c}{$q^2$ (e$^2$b$^2$)}\\
$i$&1&2&3&4&5&Exact\\
\hline
$0_1^+$ &   1.93 &   2.03 &   2.03 &   2.04 &   2.06 &2.08\\
$0_2^+$ &   2.25 &   4.71 &   5.33 &   5.36 &   5.39 &5.40\\
\hline\\
\multicolumn{7}{c}{ $\cos{3\delta}$}\\
$i$&1&2&3&4&5&Exact\\
\hline
$0_1^+$ &   0.41 &  -0.13 &  -0.15 &  -0.15 &  -0.08 &-0.08\\
$0_2^+$ &   0.38 &   1.03 &   0.51 &   0.52 &   0.54 &0.54 \\
\end{tabular}
\end{ruledtabular}
\end{table}

\section{Conclusions}
\label{sec-conclu}

Shape coexistence is a phenomenon that has become a major characteristic of atomic nuclei, all through the
nuclear mass table: ranging from the light doubly-closed shell nuclei such as $^{16}$O,$^{40}$Ca, up to heavy
nuclei with a large neutron excess such as the Sn (Z=50) and Pb (Z=82) isotopes. In almost all cases, it was
the presence of unexpectedly low-excitation energy for $0^+$ states, quite often acting as the band-head
of an intrinsic structure corresponding with a much larger collectivity as compared to the regular low-lying
states. In a number of cases, this even gave rise to highly-correlated states that ``inverted'' with the
less-correlated spherical states such as the so-called islands of inversion.

It has become clear that the presence of low-lying $0^+$ states delineates regions where different structures
sometime coexist, but depending on the proximity and their mutual coupling, give rise to important mixing
thereby sometimes masking the presence of two (or more) structures. 

In many cases, it has often been the case that the experimental energy
difference between the $0^+_1$ ground state and a low-lying intruding
$0^+_2$ state was taken as a measure of the energy difference between
energy minima associated with oblate and prolate energy surfaces in a
mean-field context. It can be concluded though, in part based on the
experimental observation of energy spectra and E2 properties for
low-lying states $0^+, 2^+, 4^+$, that such a relation is not well
founded. A first point is the fact that the total energy corresponding
with a given nuclear quadrupole shape, at the mean field level, is not
an observable. It is only after including all correlations from (i)
restoring the symmetries broken in the intrinsic frame going back to
a lab frame (making states of good angular momentum J), and, (ii)
originating from mixing the mean-fields at various deformations
(beyond mean-field calculations) that a comparison between observables
such as excitation energies, B(E2) values, ..., with theoretical
studies becomes possible.  This last step can seriously modify the
outcome at a purely mean-field level \cite{rodri00}

In the present study, we have started from a formulation which aims at
reducing the huge shell-model spaces occurring in the Pb region when
treating proton multi-particle multi-hole excitations across the Z=82
closed shell, jointly with the open neutron shell N=82-126, using the
interacting boson model symmetry-dictated approximation (IBM). This
approach keeps the essential high and low multipoles of the nuclear
effective interaction, \textit{i.e.}, the pairing and quadrupole correlations
within a boson model space.

We have analyzed in detail the even-even Hg nuclei in the region 172
$\leq$ A $\leq$ 200, in particular concentrating on the shape
coexisting phenomena observed in the 180 $\leq$ A $\leq$ 188,190
region. The IBM, including the configuration mixing between
configurations consisting of a closed proton shell at Z=82 and proton
2p-2h excitations across Z=82 (IBM-CM) has been used to describe both
energy spectra as well as E2 properties: both absolute and relative
B(E2) ratios.

The results show that, in particular at the level of the $2^+, 4^+$
states (in the interval 180 $\leq$ A $\leq$ 188), the IBM-CM is able
to correctly describe the changing mixing pattern between these two
types of configurations. This is particularly the case in order to
describe the $2^+_1$ to $2^+_2$ energy spacing and its variation as a
function of mass number A. For the higher-spin part
(J=$8^+,10^+$,...), the difference in character as intruder states
versus a regular one is evidenced when comparing energy spectra at a
given neutron number, \textit{e.g.}, N=106 for various isotones (Yb up to Hg),
as illustrated in figure 5 of ~\cite{Garc11}. We have carried out a
detailed study of the configuration mixing and the resulting wave
function content as a function of mass number A. We also point out
that studies of $\alpha$-decay hindrance factors and the mean-square
charge radii $\langle r^2 \rangle$ for the ground state, indicate a
maximal admixture of 20\% in the interval 180 $\leq$ A $\leq$ 200.

Using the coherent-state formalism, we have extracted the mean-field
energy corresponding with the IBM-CM approach (the latter formulated
in the lab system) which should correspond to the intrinsic energy
obtained with a geometric nuclear shape defined over the full
quadrupole deformation ($\beta-\gamma$ plane). This allows us to
compare with self-consistent mean-field calculations. Our conclusions
at this point were surprising. Very much in line with recent
mean-field calculations for the Hg nuclei \cite{nomura13,yao13} where
the prolate energy minimum is the deepest one for 176 $\leq$ A $\leq$
186, as compared to the oblate energy minimum (with the prolate energy
minimum disappearing above A=188), the IBM-CM results also in the prolate
energy minimum becoming the lowest one for 180 $\leq$ A $<$ 186,
being degenerate with the oblate one in A=186 and moving quickly away
for A $> 188$. These results and the observed correspondences point towards
an equivalence at the level of a mean-field description. However,
it appears that the dynamics involved in the IBM-CM are able to result in
a set of observables that are overall consistent with the large set of
experimental data obtained in the Hg nuclei.
   
We also stress the fact that starting from the experimental data, even
though direct information on deformation properties is not present in
excitation energies, B(E2) values, branching ratios, E2 matrix
elements, it is possible to construct so-called quadrupole invariants
by making appropriate sums over the reduced E2 matrix elements. These
quadrupole invariants 
$q_2$ and $q_3$ 
need a large enough set of E2 matrix elements that have
recently been extracted for the A=$182, 184, 186$ (partly for A=188)
at the Miniball set up at REX-ISOLDE. Our IBM-CM results indicate a
large difference in the value of 
$q_2$
for the
ground state $0^+_1$ state ( $\approx 2.0-1.6$ e$^2$b$^2$) as compared
to the value derived for the $0^+_2$ state ($\approx$6 $e^2b^2$), with
the experimental data showing even larger values of 
$q_3$
for the $0^+_2$ state. In contrast to the experimental data
concerning the 
$q_3$
invariant, which is a
measure of the non-axial structure of the nuclear shape for a given
nuclear eigenstate, and is indicative for having an oblate shape in
the ground state for A $=182, 184$ and A=188, our IBM-CM results are not
so conclusive.

It may be appropriate to overlook the results that have been
obtained in the study of shape coexistence within the context of a
symmetry-dictated truncation of the shell model, \textit{i.e.}, the IBM-CM
passing from the Z=82 closed shell Pb nuclei, over the Hg nuclei into
the Pt nuclei.  Whereas in the first two series of isotopes, shape
coexistence shows up clearly at the level of energy spectra,
substantiated by the large experimental efforts to disentangle finer
details in the wave functions via measurements of charge radii,
lifetime measurements, in-beam spectroscopy, etc, the Pt nuclei are
exhibiting, at first sight, a single collective structure. It looks like
the coupling between different families (0p-0h and 2p-2h excitations
across Z=82, or, spherical, oblate and prolate shapes in a mean-field
approach) in the Pb nuclei is rather moderate and only shows up
at the lower end of the two bands. The same mixing phenomenon appears
to 'disturb' the presence of two pure sets of configurations in the Hg
nuclei, this time particularly at the level of the $2^+$ states with
rather moderate mixing at the level of the $0^+$ states. The behavior
in the excitation energy of the two close-lying $2^+$ states, however,
is changing albeit in a rather smooth way, somehow concealing the
configuration mixing. This latter feature becomes dominant in the Pt
nuclei where the mixing occurs at the level of the $0^+$ state and
results in a rather low-lying excited $0^+$ state with a particular
mass dependence of its excitation energy.

\section{Acknowledgment}
 We are very grateful to Katarzyna Wrzosek-Lipska, Elisa Rapisarda and
Liam Gaffney for generous sharing of their most recent results on
Coulomb excitation (A=182-188), $\beta$-decay results for A=182 and
A=184 and new lifetime data for A=184 and A=186.  We thank M. Huyse,
P.Van Duppen for continuous interest in this research topic and
J.L.~Wood for stimulating discussions in various stages of this work
and his diligence to quote from his unpublished notes.  Financial
support from the ``FWO-Vlaanderen'' (KH and JEGR) and the InterUniversity
Attraction Poles Programme - Belgian State - Federal Office for
Scientific, Technical and Cultural Affairs (IAP Grant No.  P7/12, is
acknowledged.  This work has also been partially supported by the
Spanish Ministerio de Econom\'{\i}a y Competitividad and the European
regional development fund (FEDER) under Project No.
FIS2011-28738-C02-02, by Junta de Andaluc\'{\i}a under Project No.
FQM318, and P07-FQM-02962, and by Spanish Consolider-Ingenio 2010
(CPANCSD2007-00042).

\end{document}